\documentclass[12pt]{article}
\usepackage{graphicx,psfrag,epsf,color}
\usepackage{amsmath,amssymb,amsfonts}

\newcommand{\braket}[1]{\langle #1 \rangle}
\newcommand{\Tprod}[1]{{\mathrm T}\lbrack #1 \rbrack}

\catcode`@=11
\newcount\@tempcntc
\def\@citex[#1]#2{\if@filesw\immediate\write\@auxout{\string\citation{#2}}\fi
  \@tempcnta\z@\@tempcntb\m@ne\def\@citea{}\@cite{\@for\@citeb:=#2\do
    {\@ifundefined
       {b@\@citeb}{\@citeo\@tempcntb\m@ne\@citea\def\@citea{,}{\bf ?}\@warning
       {Citation `\@citeb' on page \thepage \space undefined}}%
    {\setbox\z@\hbox{\global\@tempcntc0\csname b@\@citeb\endcsname\relax}%
     \ifnum\@tempcntc=\z@ \@citeo\@tempcntb\m@ne
       \@citea\def\@citea{,}\hbox{\csname b@\@citeb\endcsname}%
     \else
      \advance\@tempcntb\@ne
      \ifnum\@tempcntb=\@tempcntc
      \else\advance\@tempcntb\m@ne\@citeo
      \@tempcnta\@tempcntc\@tempcntb\@tempcntc\fi\fi}}\@citeo}{#1}}
\def\@citeo{\ifnum\@tempcnta>\@tempcntb\else\@citea\def\@citea{,}%
  \ifnum\@tempcnta=\@tempcntb\the\@tempcnta\else
   {\advance\@tempcnta\@ne\ifnum\@tempcnta=\@tempcntb \else \def\@citea{--}\fi
    \advance\@tempcnta\m@ne\the\@tempcnta\@citea\the\@tempcntb}\fi\fi}
\catcode`@=12

\begin{document}
\allowdisplaybreaks

\begin{titlepage}

\begin{flushright}
TTK-10-29\\
SFB/CPP-10-26 \\
April 13, 2010
\end{flushright}

\vskip1.5cm
\begin{center}
\Large\bf\boldmath
Electroweak non-resonant NLO corrections 
to $e^+e^- \to W^+W^-b\bar{b}$ in the $t\bar{t}$ resonance region
\end{center}

\vspace{1cm}
\begin{center}
{\sc M.~Beneke}, {\sc B.~Jantzen} and {\sc P. Ruiz-Femen\'\i a}\\[5mm]
  {\it Institut f\"ur Theoretische Teilchenphysik und Kosmologie,\\
RWTH Aachen University,\\
D -- 52056 Aachen, Germany}\\[0.3cm]
\end{center}

\vspace{2cm}
\begin{abstract}
\noindent
We analyse subleading electroweak effects in the top anti-top resonance 
production region in $e^+e^-$ collisions which arise due to the decay 
of the top and anti-top quarks into the $W^+W^-b\bar{b}$ final state. 
These are NLO corrections adopting the 
non-relativistic power counting 
$v\sim \alpha_s\sim \sqrt{\alpha_{\rm EW}}$. In 
contrast to the QCD corrections which have been calculated (almost)
up to NNNLO, the parametrically larger 
NLO electroweak contributions have not been completely 
known so far, but are mandatory for the required accuracy at 
a future linear collider. The missing parts of these NLO contributions
arise from matching coefficients of non-resonant production-decay
operators in unstable-particle effective theory which correspond to 
off-shell top production and decay and other non-resonant 
irreducible background processes to $t\bar t$ production. 
We consider the total cross section of the 
$e^+e^- \to W^+W^-b\bar{b}$ process and additionally
implement cuts on the invariant masses of the $W^+b$ and $W^-\bar{b}$ 
pairs.
\end{abstract}
\end{titlepage}

\section{Introduction}

The top quark is currently known from direct production at the 
Fermilab Tevatron to weigh $m_t=173.1 \pm 0.6 \,\mbox{(stat.)} 
\pm 1.1 \,\mbox{(syst.)}\,$GeV~\cite{tev:2009ec}, and an increase 
in precision to about $1\,$GeV is 
expected soon from the Large Hadron Collider (LHC). From a 
threshold scan of the $e^+ e^-\to t\bar t$ cross section 
at the planned International Linear Collider (ILC), however, 
a precision of $30\,$MeV can be achieved 
experimentally~\cite{Martinez:2002st}. Aside from determining a 
fundamental parameter of the Standard Model, accurate top-mass
measurements constrain the quantum fluctuations from 
non-standard interactions in electroweak precision measurements. 
Other characteristics of the top quark 
such as its width and Yukawa coupling provide information 
about its coupling to other particles and the mechanism of 
electroweak symmetry breaking. For these reasons (and perhaps also 
because it represents the cleanest non-relativistic system bound 
by the colour force) top-quark pair 
production near threshold in $e^+ e^-$ annihilation has been 
thoroughly investigated following the 
non-relativistic approach 
of~\cite{Fadin:1987wz,Fadin:1988fn,Strassler:1990nw}, which 
treats the leading colour-Coulomb force exactly to all orders 
in perturbation theory. In this framework, where the strong 
coupling $\alpha_s$ is of the same order as $v$,
the small relative velocity of the top and anti-top, 
next-to-next-to-leading order (NNLO) corrections have been available 
for some
time~\cite{Hoang:1998xf,Melnikov:1998pr,Beneke:1999qg,Hoang:1999zc,Yakovlev:1998ke,Nagano:1999nw,Penin:1998mx},
next-to-leading and some higher-order logarithms of $v$ 
have been summed to all 
orders~\cite{Hoang:2000ib,Hoang:2001mm,Hoang:2003ns,Pineda:2006ri}, and 
the third-order (NNNLO) cross section is now known almost 
completely~\cite{Beneke:2005hg,Beneke:2008ec,Beneke:2008cr}, 
which requires input from three-loop matching 
coefficients~\cite{Marquard:2006qi}, 
potentials~\cite{Smirnov:2008pn,Anzai:2009tm,Smirnov:2009fh,Brambilla:1999qa,Kniehl:2002br}, 
and third-order S-wave energy levels and 
residues~\cite{Beneke:2005hg,Penin:2005eu,Beneke:2007gj,Beneke:2007pj}.
The full NNNLO result should finally clarify the question whether 
the QCD corrections can be calculated with the required precision.

Here we focus on a different issue, non-resonant production of 
the physical final state after top decay, which though known 
for some time (see, e.g., the discussion in \cite{Beneke:1999ue}) has 
been left aside up to now. The top quark is unstable with a 
significant width $\Gamma_t$ of about $1.5\,$GeV due to the electroweak 
interaction. The width is essential in threshold production, since 
it prevents the top and anti-top from forming a bound 
state~\cite{Bigi:1986jk} and causes a broad resonance structure 
in the energy dependence of the cross section on top of the 
increase due to the opening-up of the two-particle phase space. 
However, once the top width is included, due to top decay, the physical 
final state is $W^+ W^- b\bar b$ -- at least 
if we neglect the decay of top into strange and down quarks, as 
justified by $V_{tb}\approx 1$, and consider $W$ bosons as 
stable, which we may assume for the purpose of discussion in the 
present paper. The  $W^+ W^- b\bar b$ final state can be reached through 
non-resonant top production or background processes containing 
no or only single top quarks. Both effects are not included in the 
non-relativistic treatment. Adopting a standard counting scheme where 
$\alpha_{\rm EW}\sim \alpha_s^2$, we find that the leading 
non-resonant and off-shell effects are next-to-leading order (NLO) for 
the total cross section, since there is an additional power 
of $\alpha_{\rm EW}$ but no phase-space suppression, hence 
the relative correction is $\alpha_{\rm EW}/v\sim \alpha_s$. 
We note that some of the higher-order NNLO corrections due to the 
finite top width have already been calculated in~\cite{Hoang:2004tg}.

In this paper we calculate the complete NLO ``electroweak'' 
contributions to the $e^+ e^-\to W^+ W^- b\bar b$ process 
in the $t\bar t$ resonance 
region, for the total cross section as well as including 
invariant-mass cuts on the $Wb$ pairs. The calculation is performed 
with unstable-particle effective field 
theory~\cite{Beneke:2003xh,Beneke:2004km}, which provides the  
framework for consistently including resonant and non-resonant 
effects while maintaining an expansion in the small parameters 
of the problem.  There are many similarities with 
$W$-pair production considered in~\cite{Beneke:2007zg,Actis:2008rb}, though 
the top-quark case is technically more complicated due to the 
presence of a massive particle (the $W$ boson) in top decay.
We note that an alternative approach has been developed in 
parallel~\cite{Hoang:2008ud,Hoang:2010gu} that includes the effects of 
invariant-mass cuts on the $Wb$ pairs entirely through calculations 
in non-relativistic effective theory. This works if the 
invariant-mass cuts around $m_t$ are neither very loose nor very
tight, and provided that the non-resonant background processes 
are small, which must be checked by other means. The present 
approach removes both these restrictions and can be used in 
particular for the total cross section. We shall see that when 
both approaches are applicable, they agree well for 
top-quark production.

There is an interesting conceptual issue concerning the pure 
QCD calculation of the $t\bar t$ cross section. It 
exhibits an uncanceled ultraviolet divergence (here regularized 
dimensionally) 
\begin{equation}
\sigma_{t \bar t} \propto \frac{\alpha_s\Gamma_t}{\epsilon} 
\propto \frac{\alpha_s \alpha_{\rm EW}}{\epsilon}
\end{equation}  
at NNLO, which arises from the logarithmic overall divergence in the 
two-loop non-relativistic correlation function, whose imaginary part 
gives the cross section. The overall divergence is polynomial 
in the non-relativistic energy $E$ of the top quarks, but 
contributes to the cross section, since the correlation 
function is evaluated at complex values $E\to E+i\Gamma_t$. 
As discussed in \cite{Beneke:2008cr} this divergence cancels 
with an infrared divergence that appears in the non-resonant 
term in unstable-particle effective theory from the diagram 
corresponding to off-shell top-quark decay. The lesson from 
this is that the pure QCD result alone that is usually shown in 
the literature is inconsistent 
theoretically and must be embedded in the systematic 
calculation of the $e^+ e^-\to W^+ W^- b\bar b$ process. In 
the present paper, however, we are concerned with NLO 
accuracy, where no explicit divergence arises in dimensional 
regularization. But the problem at NLO is in fact worse. 
Inspection of the one-loop non-relativistic correlation function 
shows that it is linearly ultraviolet divergent, or more generally 
exhibits a cut-off dependence of the NLO order, just as the 
non-resonant contribution that we calculate below has a linear 
infrared divergence. Both require regularization and as usual 
dimensional regularization yields finite (but
regularization-dependent) values in four dimensions for 
linearly divergent integrals. The conclusion 
is that the non-resonant NLO contribution that we consider here 
is mandatory to obtain a regularization-independent result 
for top anti-top production, more precisely $W^+ W^- b\bar b$ 
production in the $t\bar t$ threshold region, at NLO.

The outline of the paper is as follows. In Section~\ref{sec:method} 
we briefly review unstable-particle effective theory. We list 
the terms that contribute at NLO to the 
$e^+ e^-\to W^+ W^- b\bar b$ process, when counting 
$\alpha_s\sim \sqrt{\alpha_{\rm EW}}\sim v$, and explain that 
the non-resonant electroweak contribution is obtained from 
the $e^+ e^- \to t \,W^- \bar b$ amplitude and its charge-conjugate 
with the top-quark width set to zero. The computation is described 
and the results are summarized in Section~\ref{sec:nonresonant} 
and Appendix~\ref{sec:appendix}. In Section~\ref{sec:invmasscuts} 
we discuss the implementation of invariant-mass cuts in the 
effective-theory calculation and validate our result by comparing 
it to the Born cross section for the  $e^+ e^- \to W^+ W^- b\bar b$ 
process computed numerically with MadGraph/MadEvent. Finally, 
in Section~\ref{sec:results} we combine the new non-resonant 
contributions with the resummed leading-order QCD result and 
the electromagnetic correction due to the electromagnetic Coulomb 
potential, which constitutes another electroweak NLO effect, and discuss the 
top anti-top line-shape as seen in the energy-dependence of 
the physical  $W^+ W^- b\bar b$ final state compared to the 
pure QCD calculation of $t\bar t$ production.
We conclude in Section~\ref{sec:conclusion}.

\section{Method of calculation}
\label{sec:method}

\subsection{Unstable-particle effective theory for pair 
production near threshold}

The cross section for the 
$e^+ e^-\to W^+ W^- b\bar b$ process (inclusive but also in the case 
where we allow for cuts in the invariant masses of the $W^+b$ and 
$W^-\bar{b}$ subsystems) is obtained from the
$W^+bW^-\bar{b}$ cuts of the $e^+e^-$ forward-scattering amplitude. 
In the energy region $\sqrt{s}\approx 2 m_t$ 
close to the top anti-top production threshold, the amplitude is 
dominated by the production of resonant top quarks with small 
virtuality. This allows us to integrate out hard modes with scale 
$m_t$ and represent the forward-scattering amplitude as the 
sum of two terms~\cite{Beneke:2003xh,Beneke:2004km}, 
\begin{eqnarray}
\label{eq:master}
i {\cal A} &=&\sum_{k,l} C^{(k)}_p  C^{(l)}_p \int d^4 x \,
\braket{e^- e^+ |
\Tprod{i {\cal O}_p^{(k)\dagger}(0)\,i{\cal O}_p^{(l)}(x)}|e^- e^+}
\nonumber\\ 
&& + \,\sum_{k} \,C_{4 e}^{(k)} 
\braket{e^- e^+|i {\cal O}_{4e}^{(k)}(0)|e^- e^+}.
\\[-0.5cm]
\nonumber
\end{eqnarray}
In the subsequent discussion of top-quark pair production near
threshold we follow closely the formalism described
in~\cite{Beneke:2007zg} for four-fermion production in $e^+ e^-$
collisions in the energy region of the $W$-pair production threshold, 
and refer to that paper for further details on the method. 

The matrix elements in (\ref{eq:master}) are evaluated in the 
``low-energy'' effective theory, which includes elements of 
soft-collinear and non-relativistic effective theory. The first term 
on the right-hand side of (\ref{eq:master}) describes the production 
of a resonant $t\bar t$ pair in terms of production (decay)
operators ${\cal O}_p^{(l)}(x)$ (${\cal O}_p^{(k)\dagger}(x)$) 
with short-distance coefficients $C^{(k,l)}_p$. The
second term accounts for the remaining non-resonant contributions. 
Note that there is no separate term for the production of
one resonant and one off-shell
top~\cite{Beneke:2007zg,Hoang:2010gu},
since such configurations are 
not sensitive to the particular low-energy dynamics that develops 
at the pair-production threshold. They are effectively short-distance
and included in the coefficient functions $C_{4 e}^{(k)}$ of  
non-resonant production-decay operators
${\cal O}_{4e}^{(k)}(0)$.

The standard pure QCD calculation of top-pair production near
threshold is entirely contained in the resonant term. In this paper 
we are mainly concerned with the calculation of the leading 
contribution to $C_{4 e}^{(k)}$, which represents an NLO correction 
to the forward-scattering amplitude ${\cal A}$, since the first term 
in  (\ref{eq:master}) is of order $\alpha_{\rm EW}^2 v$, while the 
second is ${\cal O}(\alpha_{\rm EW}^3)$. Nevertheless, for comparison 
and to discuss other NLO non-QCD effects, we briefly summarize 
the expressions relevant to the leading-order resonant term. There 
are only two production operators at this order, 
\begin{equation}
{\cal O}_p^{(v)} =
\bar{e}_{c_2} \gamma_i e_{c_1}
\,\psi_t^\dagger\sigma^i \chi_t,
\quad\qquad
{\cal O}_p^{(a)} =
\bar{e}_{c_2} \gamma_i\gamma_5 \,e_{c_1}
\,\psi_t^\dagger\sigma^i \chi_t,
\label{LPlead}
\end{equation}
where $\psi_t$ ($\chi_t$) denotes the non-relativistic top (anti-top) 
field, $e_{c}$ a collinear electron field with large, light-like 
momentum in a direction labelled by $c$, and $\gamma_i=-\gamma^i$. We drop 
the Wilson lines required to make the operators gauge-invariant, since they
do not contribute to our calculation. The tree-level 
coefficient functions of these operators are computed from the 
tree-level amplitude of the process $e^+ e^-\to t\bar t$ at production
threshold $\sqrt{s}=2 m_t$ and are given by (keeping the full $s$-dependence
in the photon and $Z$ propagators)
\begin{eqnarray}
C_p^{(v)} &=& 4\pi\alpha
\,\left[ \frac{Q_t Q_e}{s} + \frac{v_e v_t}{s-M_Z^2} \,
\right],
\nonumber\\
C_p^{(a)} &=& -4\pi\alpha
\,\frac{a_e v_t}{s-M_Z^2} \,,
\end{eqnarray}
where $\alpha$
denotes the electromagnetic coupling, $Q_f$ the electric charge 
of fermion species $f$ in units of the positron charge ($Q_t=2/3$, $Q_e=-1$),
and $M_Z$ the $Z$-boson mass. The vector and
axial-vector coupling of the fermion $f$ to the $Z$-boson are given by
\begin{eqnarray}
&& v_f =\frac{T_3^f-2Q_f s_w^2}{2s_w c_w},
\quad\qquad
a_f=\frac{T_3^f}{2s_w c_w},
\end{eqnarray}
where $s_w$ ($c_w$) is the sine (cosine) of the weak mixing angle
and $T_3^f$ the third component of
the weak isospin of the fermion. In terms of these quantities the 
leading-order cross section is given by
\begin{equation}
\sigma^{(0)}_{t\bar t} = \frac{1}{s}\,\mbox{Im}\,{\cal A}^{(0)}
= \left[{C_p^{(v)}}^2+{C_p^{(a)}}^2\right] 2N_c 
\,\mbox{Im} \,G_{\rm C}^{(0)}(0,0;{\cal E}),
\label{eq:sigmaLO}
\end{equation}
where $E\equiv \sqrt{s}-2 m_t$ and ${\cal E}\equiv E+i\Gamma_t$, 
$N_c=3$ equals the number of colours, 
and $ G_{\rm C}^{(0)}(0,0;{\cal E})$  denotes the 
$\overline{\rm MS}$-renormalized zero-distance Coulomb Green 
function~\cite{Beneke:1999zr} 
  \begin{eqnarray}
  \label{coulombGF}
    G_{\rm C}^{(0)}(0,0;{\cal E}) &=&  -\frac{m_t^2}{4\pi} \,\Bigg\{
    \sqrt{-\frac{{\cal E}}{m_t}} + \alpha_s C_F \bigg[
    \frac{1}{2}\ln \bigg(\! -\!\frac{4\,m_t {\cal
        E}}{\mu^2}\bigg)-\frac{1}{2} \nonumber\\
 && +\,\gamma_E
    +\psi\bigg(1-\frac{\alpha_s C_F}{2\sqrt{-{\cal E} \slash m_t}}\bigg)
    \bigg]\!\Bigg\}.
  \end{eqnarray}
Here $\gamma_E$ is the Euler-Mascheroni constant, 
$\mu$ the scale introduced in dimensional regularization,
and $\psi$ the Euler psi-function.
(The dependence on $\mu$ and the regularization drops out in
$\mbox{Im} \,G_{\rm C}^{(0)}(0,0;{\cal E})$.)

The leading contribution from non-resonant production-decay
operators ${\cal O}^{(k)}_{4e}$ to (\ref{eq:master}) arises from
four-electron operators of the form
\begin{equation}
\label{eq:4eLag}
{\cal O}^{(k)}_{4e}=
\bar e_{c_1}\Gamma_1 e_{c_2} \,\bar e_{c_2}\Gamma_2 e_{c_1},
\end{equation}
where $\Gamma_1$, $\Gamma_2$ are Dirac matrices. The calculation of 
the short-distance coefficients
$C^{(k)}_{4e}$ is performed in standard fixed-order perturbation
theory in the full electroweak theory. In particular, the 
top propagator is the free propagator not including the top width, 
which ensures that the amplitude depends only on the short-distance 
scales. Self-energy insertions are treated perturbatively and 
are not resummed into the propagators.
The leading contribution to the forward-scattering amplitude
arises from the one-loop diagrams with two top propagators. 
Just as in the case of $W$-boson pair production~\cite{Beneke:2007zg} 
the imaginary part of these diagrams vanishes in dimensional 
regularization to all orders in the expansion in 
$E=\sqrt{s}-2 m_t$ in the hard region. Thus the leading 
imaginary parts of $C^{(k)}_{4e}$ arise from two-loop
diagrams of order $\alpha_{\rm EW}^3$ shown in Figure~\ref{fig1} 
below. The corresponding contribution to the cross section is
\begin{equation}
\label{eq:nonres}
\sigma_{\text{non-res}} = \frac{1}{s}\,
\sum_k \,\mbox{Im}\left[C_{4 e}^{(k)}\right]\,
\braket{e^- e^+|i {\cal O}_{4e}^{(k)}(0)|e^- e^+}.
\end{equation}
It is convenient not to calculate the $C_{4 e}^{(k)}$ separately, 
but directly the sum (\ref{eq:nonres}) when one considers the 
unpolarized cross section.  
Technically, this simply amounts to the calculation of the 
spin-averaged tree-level processes $e^+ e^- \to t W^- \bar b$ and 
$e^+ e^- \to \bar t W^+ b$ with no width supplied to the 
intermediate top-quark propagators. Instead, the divergence 
from the top-quark propagators going on-shell is regularized
dimensionally.

\subsection{NLO corrections related to the top-quark instability and 
electroweak interactions}
\label{NLOsummary}

In the following we list the NLO corrections to the  
$e^+ e^-\to W^+ W^- b\bar b$ process in the top anti-top resonance 
region $\sqrt{s}\approx 2 m_t$. We do not discuss in this paper 
the pure QCD NLO corrections consisting 
of the one-loop QCD correction to the $\gamma^* (Z) \,t\bar t$ vertex 
and the one-loop correction to the QCD Coulomb potential, since 
they are standard and included in the pure QCD calculations  
in the literature. Instead we focus on NLO corrections which 
arise once we consider the $W^+ W^- b\bar b$ final state 
(as required by the top instability), and other electroweak 
effects:

\begin{itemize} 
\item The single insertion of the (leading-order) electromagnetic 
Coulomb interaction instead of the colour Coulomb interaction 
is an NLO effect, since $\alpha/\alpha_s\sim \alpha_s$. This  
is trivially accounted for, including some higher-order terms, by replacing 
$\alpha_s C_F \to \alpha_s C_F+\alpha Q_t^2$ in (\ref{coulombGF}). 
This effect has already been studied in \cite{Pineda:2006ri,Hoang:2010gu}.
  
\item The one-loop QCD correction to the on-shell top 
width \cite{Jezabek:1988iv} is also 
trivially included by using the one-loop corrected expression for 
$\Gamma_t$ in the variable ${\cal E}\equiv E+i\Gamma_t$ 
in (\ref{coulombGF}). We do not discuss this effect further in this
paper, however, 
since in the pure QCD calculation the top width is considered 
as an adjustable input parameter, and thus may be thought of as 
representing the one-loop corrected expression.
 
\item Corrections coming from gluon exchange involving the bottom
quarks in the final state (QCD interference effects) appear at 
NLO but vanish at NLO for the total cross section
\cite{Melnikov:1993np}. We show in Section~\ref{sec:invmasscuts} 
that they are parametrically suppressed in the presence of 
invariant-mass cuts, provided the cuts are loose in the sense 
defined below (see also \cite{Hoang:2010gu}). Since we focus on 
loose cuts in the final results, no calculation of interference 
effects is required.

\item Non-resonant contributions start at NLO, as discussed above. 
They arise from the hard contributions to the Born cross section for 
$e^+e^-\to W^+ W^- b\bar{b}$, which determine the matching
coefficients of the four-electron 
production-decay operators. We calculate this contribution in 
Section~\ref{sec:nonresonant}.
\end{itemize}
We should note that any three-particle final state produced 
through the electroweak interaction in $e^+ e^-$ annihilation 
contributes at NLO  (namely ${\cal O}(\alpha_{\rm EW}^3)$) 
provided the  $W^+ W^- b\bar b$ final state can be reached through 
the resonant decay of one of the three particles with a 
significant branching fraction of the decaying resonance.
The obvious relevant cases are $t \,W^-\bar{b}$ and $\bar{t} \,W^+ b$ 
considered in Section~\ref{sec:nonresonant}; others are 
$e^+ e^- \to W^+ W^- H$ for low Higgs masses, when $H\to b\bar b$ is 
the dominant decay mode, and  $e^+ e^- \to H b\bar b$, when 
the Higgs decays resonantly to $W^+W^-$ with a large branching 
fraction. The process $e^+ e^- \to W^+ W^- Z$ with $Z$ decaying 
to $b\bar b$ is less prominent, since the $Z\to b\bar b$ branching 
fraction is only about 15\%. We do not compute these additional 
three-particle processes in the present paper, since they 
are further suppressed when invariant-mass cuts in the top anti-top 
signal region are applied. Moreover, in a  realistic analysis 
these contributions would be eliminated anyway by invariant-mass cuts on 
the resonance (Higgs or $Z$) decay products.

\section{Computation of the four-electron matching coefficients at NLO}
\label{sec:nonresonant}

The four-electron matching coefficients $C_{4e}^{(k)}$ originate 
from the hard contributions of the $e^+e^-$ forward-scattering
amplitude. The hard momentum region expansion dictates that the 
top-quark self-energy insertions are treated perturbatively, 
since the top lines are formally far off-shell, 
$p_t^2-m_t^2 \sim {\cal O}(m_t^2) \gg \Sigma(p_t^2)$. Accordingly the 
calculation of the coefficients $C_{4e}^{(k)}$ is performed in 
fixed-order perturbation theory in the full electroweak theory 
with no resummation of self-energy insertions in the top-quark 
propagator~\cite{Beneke:2007zg} and supplemented with an expansion 
of the amplitudes in $\delta=s/(4m_t^2)-1 \sim v^2$. The corresponding
contributions to the imaginary part of the forward-scattering
amplitude are then reproduced by those imaginary parts of the 
short-distance coefficients $C_{4e}^{(k)}$ which can be identified 
with the $W^+W^- b\bar{b}$ final state. The extraction of these 
imaginary parts is more conveniently performed by evaluating the 
relevant cuts of the hard $e^+e^-$ forward-scattering diagrams
directly. The leading contribution arises from the cut one-loop 
diagram with a top and anti-top propagator; however, the hard region 
of this diagram vanishes in dimensional regularization to all orders
in the $\delta$-expansion.

\begin{figure}[t]
\begin{center}
\includegraphics[width=0.95\textwidth]{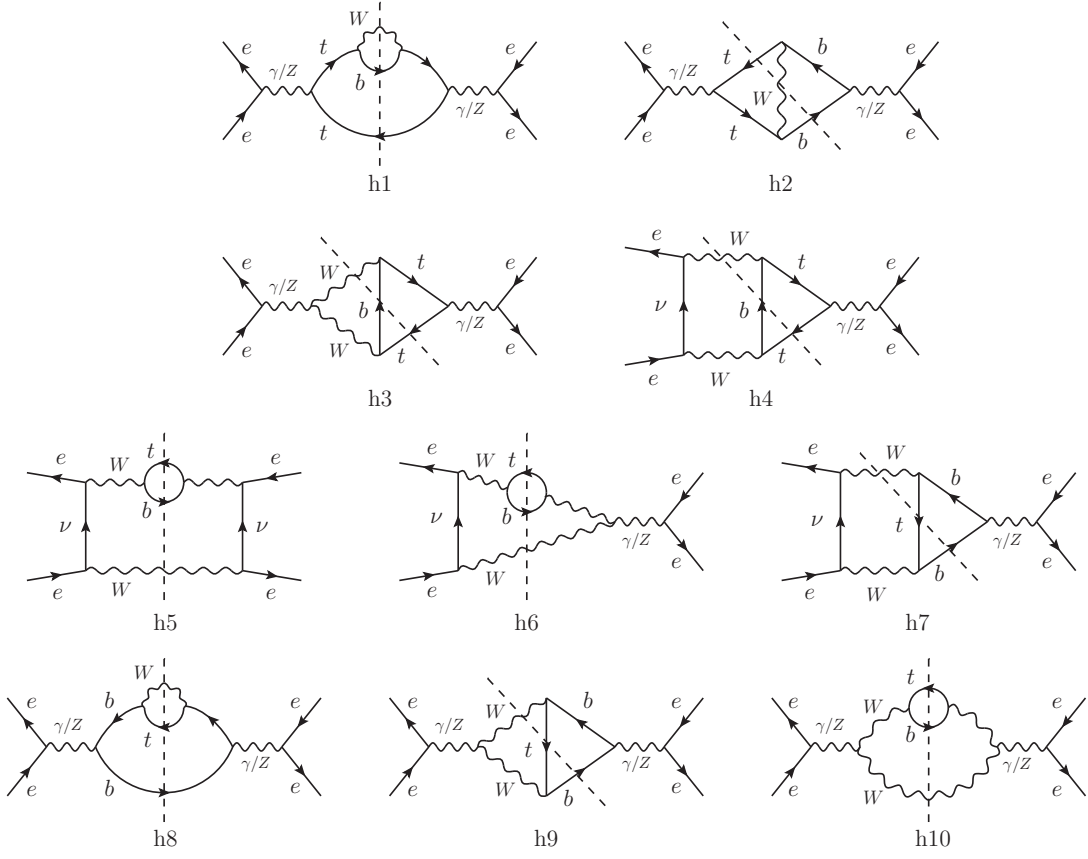}
\caption{Two-loop forward-scattering amplitude diagrams with 
$\bar{t} b W^+$ cuts. $t \bar{b} W^-$ cuts and symmetric diagrams 
are not shown.}
\label{fig1}
\end{center}
\end{figure}

A non-zero imaginary contribution to the short-distance coefficients 
$C_{4e}^{(k)}$ arises, in the unitary gauge, from the cut two-loop 
diagrams shown in Figure~\ref{fig1} expanded near threshold 
(in $\delta=s/(4m_t^2)-1$). Including cuts of the top lines 
is consistent with the expansion of the imaginary part of the resummed
top-quark propagator in the hard 
region~\cite{Beneke:2007zg}\footnote{Recall that applying the 
Cutkosky rule for an internal line leads to the replacement of the 
propagator by its imaginary part. The leading term obtained by the 
expansion of the imaginary part in the hard region leads to a Dirac 
delta function, which is interpreted as if a top line with no 
self-energy insertion is cut.}.
The contribution to the $W^+W^-b\bar{b}$ cross section from diagrams 
$h_1$--$h_{10}$ can be interpreted as the $\bar{b} W^-$ pair
originating from a nearly on-shell anti-top decay, while the $b W^+$ pair
is produced non-resonantly, either from a highly virtual top 
(diagrams $h_1$--$h_4$), or without an intermediate top as in the
truly single-resonant diagrams $h_5$--$h_{10}$. The contributions from
$t \,\bar{b} W^-$  cuts, which are not displayed
in Figure~\ref{fig1}, yield identical results, by virtue of 
$CP$-invariance. The calculation of the amplitudes in
Figure~\ref{fig1} amounts to the calculation of the squared and 
phase-space integrated matrix element of the on-shell processes 
$e^+e^-\to t \,W^-\bar{b}$ and $e^+e^- \to \bar{t} \,W^+ b$  in ordinary 
perturbation theory with no width added to top propagators and 
divergences regularized dimensionally for consistency with the
calculation of the resonant contribution to (\ref{eq:master}).
The result is parametrically of order $\alpha_{\rm EW}^3$, and 
hence represents an NLO correction relative to the 
${\cal O}(\alpha_{\rm EW}^2 v)$ leading-order cross section.

Beyond the NLO non-resonant terms that we calculate in this paper, 
the NNLO contributions are given by the 3-loop cut diagrams containing
the $\alpha_s$-corrections to the amplitudes of Figure~\ref{fig1}, 
while the next order in the $\delta$-expansion of the two-loop 
diagrams contributes only at N$^3$LO. It is also interesting to note  
that pure background diagrams, {\it i.e.} diagrams without any top
lines, start to contribute to the imaginary parts
of the short-distance coefficients $C_{4e}^{(k)}$ at N$^3$LO through
the $W^+W^- b\bar{b}$ cuts of pure EW 3-loop diagrams. An exception 
to this are the three-particle final states with resonant decay of one
of the particles to $W^+W^- b\bar{b}$ with large branching fraction 
as discussed in the previous section.

Turning now to the calculation, it proves convenient to cast the 
contributions from diagrams $h_1$--$h_{10}$ (at order $\delta^0$
and for $m_b=0$) in the form 
\begin{equation}
\int_{\Delta^2}^{m_t^2} dp_t^2 \,(m_t^2-p_t^2)^{\frac{d-3}{2}} 
H_i\Big(\frac{p_t^2}{m_t^2},\frac{M_W^2}{m_t^2} \Big)\,
\label{eq:Hform}
\end{equation}
where we have introduced the variable
\begin{equation}
p_t^2 \, =\,(p_b+p_{W^+})^2
\,,
\label{eq:pt}
\end{equation}
which corresponds to the invariant mass of the $bW^+$ system formed 
by the $b$ and $W^+$ lines that are cut. For diagrams $h_1$--$h_4$, 
where the $bW^+$ system comes from the decay of the top line, the 
momentum $p_t$ corresponds to the momentum of the top line.
The upper limit on the $p_t^2$ integration in (\ref{eq:Hform})
corresponds to the kinematic limit associated to
the $\bar{t}bW^+$ final state when $\delta\to 0$. The kinematic lower 
limit is given by $\Delta^2=M_W^2$. The case of loose cuts on the 
$bW^+$ and $\bar{b}W^-$ invariant masses that are discussed
in Section~\ref{sec:invmasscuts} can be incorporated easily in the 
computation of the four-electron matching coefficients by setting the 
lower bound of the $p_t^2$ integration to $\Delta^2=m_t^2-\Lambda^2$, 
with the parameter $\Lambda\equiv \Lambda_-$ defined 
in~(\ref{eq:lambda}) below.
The factor $(m_t^2-p_t^2)^{(d-3)/2}$ in (\ref{eq:Hform}) regularizes 
the endpoint singularity in diagram $h_1$ at $p_t^2=m_t^2$ due to the 
two top propagators that make $H_1$ behave as 
\begin{equation}
H_1\Big(\frac{p_t^2}{m_t^2},\frac{M_W^2}{m_t^2} \Big) 
\;\;\stackrel{p_t^2\to m_t^2}{\rightarrow} \;\;\mbox{const} \times 
\frac{1}{(m_t^2-p_t^2)^2}\,.
\end{equation}
The use of dimensional regularization within the threshold expansion
thus provides a natural way to regularize the phase-space
singularities in the hard region where the top propagators have no
width. Dimensional regularization {\em must} be used for consistency, 
when the pure QCD 
calculation is performed with the dimensionally regularized 
non-relativistic effective theory. In the calculation of the other 
diagrams $h_2$--$h_{10}$ we can set $d=4$, since they contain at 
most one internal top propagator leading to an integrable square-root 
endpoint singularity in~(\ref{eq:Hform}).

The result of the calculation can be written as
\begin{align}
 \sigma^{(1)}_{\text{non-res}}  = & \,\,\frac{32 \pi^2 \alpha^2}{s}\,
\frac{\Gamma_t^{\rm Born}}{m_t}
\bigg\{ 
\Big[ Q_t^2 C_{\gamma\gamma}(s)  -2 v_t Q_t  C_{\gamma Z} (s)+ 
C_{Z Z} (s) v_t^2\Big]
\frac{6\sqrt{2}}{\pi^2} \frac{m_t}{\Lambda}
\nonumber \\
&+ C_{\gamma\gamma} (s)
\left[  Q_t^2 h_1^{V} + 2 Q_b Q_t h_2^{V}
+ 2 Q_t h_3^V + Q_b^2 h_8 + 2 Q_b h_9 + h_{10}   \right]
\nonumber \\ 
& + C_{\gamma Z} (s)
\Big[ -2 Q_t v_t h_1^{V} + 2 Q_t a_t h_1^{VA}
-2 \big( Q_b v_t +Q_t (v_b+a_b) \big) h_2^{V}  
\nonumber \\ 
& 
\hspace{2cm}- 2 a_t Q_b h_2^{A}  -2 v_t h_3^{V} + 2 a_t h_3^{A} 
-2 \frac{c_w}{s_w} Q_t h_3^{V} 
\nonumber \\
&
\hspace{2cm}- 2  Q_b (v_b+a_b) h_8  
-2\Big(v_b+a_b+Q_b \frac{c_w}{s_w}\Big)h_9 - 2 \frac{c_w}{s_w}h_{10} \Big]
\nonumber \\ 
&
+ C_{Z Z} (s)
\Big[  v_t^2 h_1^{V} + a_t^2 h_1^{A}
-2 v_t a_t h_1^{VA} + 2 (v_b+a_b) ( v_t h_2^{V} + a_t h_2^{A} )   
\nonumber \\ 
&\hspace{1cm}
 + 2 \frac{c_w}{s_w} (v_t h_3^{V} - a_t h_3^{A}) + (v_b+a_b)^2 h_8 
 + 2 \frac{c_w}{s_w} (v_b+a_b)h_9
 + \frac{c_w^2}{s_w^2} h_{10} \Big] 
\nonumber \\
&
+ C_\gamma(s) \Big[ Q_t h_4^{V}+ h_6 + Q_b h_7 \Big]
\nonumber \\
&
+ C_Z(s) \Big[ v_t h_4^{V} + a_t h_4^{A} + \frac{c_w}{s_w} h_6 + 
(v_b+a_b) h_7 \Big]
+ \frac{1}{s_w^4} h_5 
\bigg\}\,,
\label{eq:NLOmatching}
\end{align}
where the functions 
\begin{equation}
C_{\gamma\gamma}(s) = -  Q_e^2 \, \frac{m_t^2}{4s}\;,\qquad
C_{\gamma Z}(s) =  \frac{Q_e v_e  m_t^2}{4 (s-M_Z^2)}\;,\qquad\;
C_{ZZ}(s) = - \frac{ s\,(v_e^2+a_e^2)  m_t^2}{4(s-M_Z^2)^2}\;, 
\nonumber
\end{equation}
\begin{equation}
C_\gamma(s)= \frac{Q_e m_t^2}{s_w^2 s} \;,\qquad 
C_Z(s) = \frac{(v_e + a_e)m_t^2}{s_w^2(s-M_Z^2)}
\label{eq:Cs}
\end{equation}
retain the exact $s$-dependence from the photon and $Z$-boson
propagators, and
\begin{equation}
\Gamma_t^{\rm Born}=\frac{\alpha|V_{tb}|^2 m_t}{16s_w^2} 
\frac{(1-x)^2(1+2x)}{x} 
\label{eq:Gammatop}
\end{equation}
with  $x=M_W^2/m_t^2$ is the tree-level top decay width obtained
from $t\to b W^+$ with the bottom-quark mass set to zero.
The dimensionless functions 
\begin{equation}
h_i^{X} \equiv h_i^{X}(x,\Delta^2/m_t^2) 
\label{eq:hi}
\end{equation}
with $X=V,A,VA$, are obtained from diagrams $i=1,...10$, and depend
only on the ratios\footnote{Here 
we already allow for an invariant-mass cut on $p_t^2$ as discussed 
in Section~\ref{sec:invmasscuts}. With no cut, $\Delta = M_W$ and 
$\Lambda = \sqrt{m_t^2-M_W^2}$ in (\ref{eq:NLOmatching}).}  
$(M_W/m_t)^2$ and $(\Delta/m_t)^2$ but not on the centre-of-mass 
energy $\sqrt{s}$.
To define the function $h_1^{V}$ we subtracted from 
the integrand of diagram~$h_1$ its leading singular behaviour as 
$p_t^2 \to m_t^2$ (see (\ref{eq:h1int}) in
Appendix~\ref{sec:appendix}). Adding back the subtracted expression
yields the terms in the first line of (\ref{eq:NLOmatching}) upon
$p_t^2$ integration. Integral representations of the functions $h_i^X$
which can easily be evaluated numerically
are provided in the Appendix. 

At NLO we could put $s=4m_t^2$ in the $C_i(s)$ functions appearing 
in~(\ref{eq:Cs}) and in the overall $1/s$ factor; keeping the
exact energy dependence slightly improves the quality of the
effective field theory (EFT)
expansion. Since this energy dependence
is mild compared to that of the resonant contributions 
within the range of energies relevant for the threshold region,
the non-resonant contribution $\sigma^{(1)}_{\text{non-res}}$ results 
in an almost constant shift of the cross section. 

\begin{figure}[t]
  \begin{center}
  \includegraphics[width=0.75\textwidth]{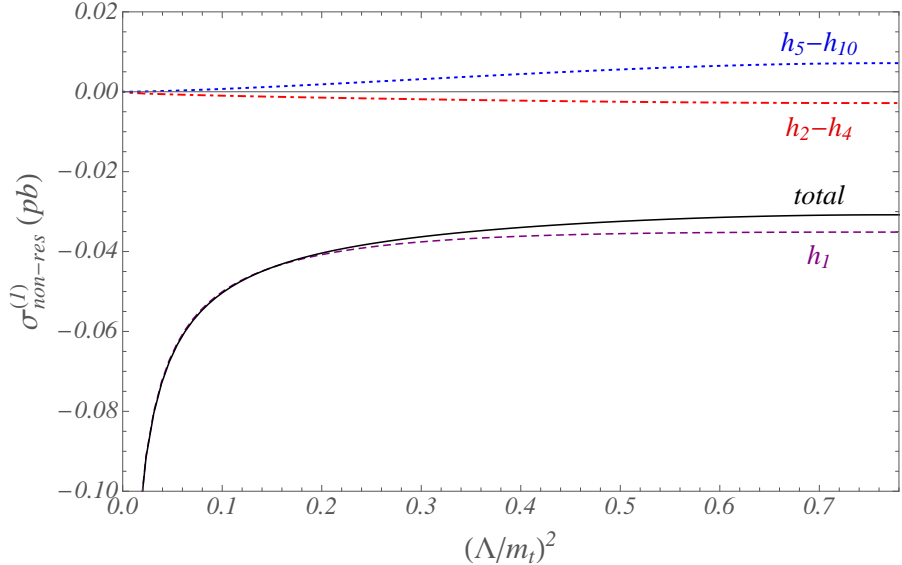}
  \caption{NLO non-resonant contributions to the cross sections 
computed at $s=4m_t^2$  as a function
of $(\Lambda/m_t)^2$, and with values
for the input parameters given in~(\ref{eq:input}). The maximal 
value of $(\Lambda/m_t)^2$ corresponding to the total cross section 
equals $1-M_W^2/m_t^2 = 0.7815$. The solid (black) line is the
total contribution $\sigma^{(1)}_{\text{non-res}}$ in (\ref{eq:NLOmatching}), and
the contributions from the diagram $h_1$, from the sum of diagrams $h_2$--$h_4$
and from the truly single-resonant diagrams $h_5$--$h_{10}$ are
shown by the dashed (purple), dash-dotted (red) and dotted (blue) lines,
respectively.
  }
  \label{fig2}
  \end{center}
\end{figure}

The size of the NLO non-resonant contributions to the cross section 
evaluated at  $s=4m_t^2$ is shown in Figure~\ref{fig2} as a function 
of $\Lambda^2/m_t^2$.  In this Figure the total cross section  
corresponds to the maximal value of 
$\Lambda^2/m_t^2 = 1-M_W^2/m_t^2 = 0.7815$. The analysis
reveals that the total (solid line in Figure~\ref{fig2}) is 
negative\footnote{The non-resonant cross section can be negative,
  since it is defined by analytic continuation to four dimensions 
of an expression that is divergent without regularization. Only the 
sum of resonant and non-resonant terms in (\ref{eq:master}) represents
a physical quantity.}
and follows closely the contribution from diagram $h_1$ alone (dashed
line) for all values of $\Lambda^2$. The sum of
the contributions of diagrams $h_2$--$h_4$ (dash-dotted line),
where the $b W^+$ is produced from a highly virtual top quark, 
is rather small, less than 3~fb for all values of $\Lambda^2$. The 
individual contributions from diagrams $h_3$ and $h_4$ are comparable 
in size with that of $h_1$, but they cancel to a large extent.
The sum of the truly single-resonant diagrams $h_5$--$h_{10}$
(dotted line in Figure~\ref{fig2}) is positive and also small,
reaching at most 7~fb for the total cross section ({\it i.e.} 
when $\Lambda^2=m_t^2-M_W^2$). 
The smallness of the single-resonant contributions is
due to large cancellations among diagrams $h_5$, $h_6$, and $h_{10}$
which individually become more important than diagram $h_1$ for large values
of $\Lambda^2$.

For easy use we provide a compact fit formula for 
the NLO non-resonant contributions to the total cross section 
(corresponding to $\Lambda^2/m_t^2=1-x$) in a reasonable range of 
top-mass values. In the range $m_t = 160$--180~GeV,
a quadratic fit to the functions $h_i^{X}(x,x)$,
\begin{equation}
h_i^{X,\rm fit}(x,x)=a + b \, \delta_{m_t}+ c\, \delta_{m_t}^2 \quad,\quad
\delta_{m_t}=\frac{m_t}{\text{170 GeV}} - 1\,,
\label{eq:hiapprox}
\end{equation}
is accurate to a precision better than
one per mille. The coefficients $a,b,c$ are given in Table~\ref{table1}.
When the functions $h_i^{X,\rm fit}(x,x)$ are combined into the
total non-resonant contribution $\sigma^{(1)}_{\text{non-res}}$ 
according to~(\ref{eq:NLOmatching}), the precision of the quadratic
fit compared to an exact (numerical) evaluation is
better than $5\cdot 10^{-5}$ for $160~\text{GeV} \le m_t \le
180~\text{GeV}$. Omitting the quadratic term in
$h_i^{X,\rm fit}(x,x)$, \emph{i.e.} setting the $c$-coefficients to zero,
the precision of $\sigma^{(1)}_{\text{non-res}}$ is still about one 
per mille or better.

\begin{table}[t]
\centering
\begin{tabular}{l c c c  }
\hline
 & $a$ & $b$ & $c$  \\
\hline
$h_1^V$ & 1.292 & -0.4022 & 1.024 \\
$h_1^A$ & -0.1631 & -0.06325 & 0.1056 \\
$h_1^{VA}$ & -0.1286 & -0.05451 & 0.08804 \\
$h_2^V$ & 0.009852 & -0.05238 & -0.005537 \\
$h_2^A$ & 0.001222 & -0.001931 & -0.005601 \\
$h_3^V$ & -0.4132 & -0.8710 & -0.2679 \\
$h_3^{A}$ & -0.04729 & -0.1324 & -0.05029 \\
$h_4^{V}$ & 0.05962 & 0.08954 & 0.02671 \\
$h_4^{A}$ & -0.006691 & -0.01716 & -0.005359 \\
$h_5$ & 0.005727 & 0.01925 & 0.02413  \\
$h_6$ & 0.07369 & 0.2965 & 0.3855 \\
$h_7$ & 0.001891 & 0.006638 & 0.004723 \\
$h_8$ & -0.007086 & -0.01070 & 0.009648 \\
$h_9$ & -0.02321 & -0.06746 & -0.02936 \\
$h_{10}$ & -0.5453 & -2.282 & -3.081 \\
\hline
\end{tabular} 
\caption{Coefficients for the fit functions 
$h_i^{X,\rm fit}(x,x)$.}
\label{table1}
\end{table}

\section{Invariant-mass cuts in the effective theory}
\label{sec:invmasscuts}

The  measurement of the $t\bar{t}$ threshold shape under realistic 
experimental conditions will require to apply cuts in the kinematic 
variables of the observed particles in order to select
$W^+W^-b\bar{b}$ events. For a general treatment of such
phase-space cuts the EFT framework will have to be modified, and 
the bottom-quark and $W$-boson degrees of freedom and eventually those
of their decay products have to be included.
However, cuts on kinematic variables that are not sensitive to the 
angular or momentum distributions of the top decay products, such as 
cuts on the invariant masses of the $bW^+$ and $\bar{b}W^-$
subsystems, can be implemented in the unstable-particle
effective-theory calculation, as discussed 
in~\cite{Actis:2008rb,Hoang:2008ud,Hoang:2010gu}.

We consider symmetric cuts on the invariant masses of the top ($M_t$) 
and anti-top ($M_{\bar{t}}$) decay products of the form
\begin{align}
m_t -\Delta M_t &  \le M_{t,\bar{t}} \le m_t +\Delta M_t\,,
\label{eq:DeltaM}
\end{align}
or equivalently 
\begin{align}
-\Lambda_-^2 &  \le M_{t,\bar{t}}^2-m_t^2 \le \Lambda_+^2\,,
\label{eq:lambda}
\end{align}
with $\Lambda_\mp^2=2m_t \Delta M_t\mp \Delta M_t^2$.
Invariant-mass cuts on the top decay products translate
to cuts on the momenta of the top ($p_t$) and anti-top ($p_{\bar{t}}$) 
circulating in the potential and hard loops
of the $e^+e^-$ forward-scattering amplitude. 
The cuts of the form~(\ref{eq:lambda}) can be implemented by inserting
a product of step-functions 
$\theta(\Lambda_+^2-[p_t^2-m_t^2])\theta(p_t^2-m_t^2 + \Lambda_-^2)$, 
and similarly for $p_{\bar{t}}^2$, in the cut loop integral. In the 
effective theory the same step functions have to be introduced into
the loops in the matrix elements of the resonant contributions, which 
reproduce the potential region of the full-theory diagrams. 
In the hard region, these step
functions modify the calculation of the matching coefficients of the 
four-electron operators, which then become dependent on the 
invariant-mass cuts $\Lambda_\mp^2$. At NLO this dependence enters 
explicitly through the lower limit in the integration that defines the
$h_i^X$ functions, see~(\ref{eq:Hform}), 
(\ref{eq:hi}). At this order, no dependence on the upper
invariant-mass cut $\Lambda_+^2$ appears because the phase-space
integration of the $\bar{t}bW^+$ final state in the hard diagrams of 
Figure~\ref{fig1} sets an upper kinematic limit
$p_{t}^2 \le m_t^2$; hence we define $\Lambda\equiv \Lambda_-$.

The two limiting cases for the invariant-mass cut corresponding to a 
loose cut, $\Lambda_\mp^2 \sim m_t^2$ (corresponding to 
$\Delta M_t\sim m_t$), and a tight cut, 
$\Lambda_\mp^2 \sim m_t \Gamma_t$ (corresponding to $\Delta M_t\sim
\Gamma_t$), yield particularly simple prescriptions
for implementing invariant-mass cuts in unstable-particle
effective theory, as it has been discussed in the context of 
$W$-pair production near threshold~\cite{Actis:2008rb}. For tight cuts
the matching coefficients of the four-electron operators
vanish because in the hard region we have  
$|p_t^2-m_t^2| \gg \Lambda_\mp^2$. Then the expansion
of the product of step functions in small variables provides  
the factor $\theta(-|p_t^2-m_t^2|)=0$. Therefore four-electron
operators do not contribute to the cross section for tight invariant-mass 
cuts. The same does not apply for the contributions from the resonant 
part, because there $p_t^2-m_t^2=2m_t r_0-\vec{r}^{\,2} \sim
\Lambda_\mp^2$, since the residual momentum
$r=(r_0,\vec{r}\,)$ has the characteristic scaling of the potential
region $r\sim (m_t v^2,m_t v)$. 
The situation is reversed if we consider loose cuts. In this case 
the implementation of invariant-mass cuts on the loop integrals from 
the effective theory has no effect at the leading order 
because we can drop the momentum dependence from the step function 
following the assumption $\Lambda_\mp^2\gg m_t r_0,\vec{r}^{\,2}\sim
(m_t v)^2$, and we are left with
$\theta(\Lambda_-^2)\theta(\Lambda_+^2)=1$. In particular, this means 
that the NLO QCD interference contributions are not affected by loose
cuts and thus vanish, as it happens for the total cross section. The 
latter is confirmed by the calculation of the QCD interference
effect with an invariant mass cut 
satisfying $\Lambda_\mp^2\gg m_t \Gamma_t$~\cite{Hoang:2010gu}, which 
yields a correction of order $m_t\Gamma_t^2/\Lambda_\mp^3$ relative 
to the LO resonant term, which is indeed parametrically suppressed 
relative to a NLO correction of order $\sqrt{\alpha_{\rm EW}}$ when 
$\Lambda_\mp \sim m_t$.
However, as already mentioned, the loose cut needs to be taken into 
account in the calculation of the matching coefficients of the 
four-electron operators as done in Section~\ref{sec:nonresonant}.

It is interesting to compare the above procedure for the treatment of 
invariant-mass cuts in the unstable-particle effective theory with 
the {\it phase-space matching} approach introduced in the recent 
paper~\cite{Hoang:2010gu}. In the latter, the matching coefficients of
the four-electron operators (called  {\it phase-space matching} 
conditions) are determined in the following way. First, cuts of the 
form (\ref{eq:DeltaM}) are introduced on the top and anti-top 
momenta in the NRQCD loop diagrams, the latter corresponding to the 
resonant contributions in~(\ref{eq:master}). Then the resulting
cut loop integrals are expanded assuming that 
$\Lambda^2 \gg (m_t v)^2, m_t \Gamma_t$, which is formally correct
as long as $\Lambda^2 \ll m_t^2$, since NRQCD describes the 
non-relativistic $t\bar{t}$ configurations and $\Lambda^2$ 
acts effectively as a cut on the non-relativistic momentum
$\vec{r}^{\,2}$. The $\Lambda$-dependent
terms that result from this expansion yield a series of the form 
$\Gamma_t/\Lambda \times \sum_{n,m} [ (m_t \Gamma_t/\Lambda^2)^n 
\times (\Lambda^2/m_t^2)^m ]$, with $n,m=0,1\dots$.
Powers of $\Lambda^2/m_t^2$ enter first from cut diagrams with 
NNLO relativistic corrections that come with a factor 
$\vec{r}^{\,2}/m_t^2$, and are a signal of the breakdown of
the series for $\Lambda\sim m_t$. The whole computation is equivalent 
to the non-relativistic expansion of the full-theory squared matrix 
elements containing the double-resonant diagrams for 
$e^+e^-\to t\bar{t}\to W^+W^-b\bar{b}$ and their interference with 
the diagrams for $e^+e^-\to W^+W^- b\bar{b}$ having only either the 
top or the anti-top in intermediate stages. Contrary to the method 
presented in the present paper, the non-resonant contributions to the
full-theory matrix element coming from the square of single-top and 
pure background diagrams (the so-called remainder contributions 
in~\cite{Hoang:2010gu}) cannot be determined within the phase-space 
matching approach. The path taken by the authors of
\cite{Hoang:2010gu} is to compute the remainder contributions by 
comparing the NRQCD cross section with the relativistic Born cross 
section for $e^+e^-\to W^+W^- b\bar{b}$ as obtained with 
MadGraph/MadEvent~\cite{Alwall:2007st}. The comparison yielded that 
the remainder terms are below 5~fb for invariant-mass cuts 
$\Delta M_t\le 35$~GeV, and hence were neglected given the theoretical 
precision goal aimed at in~\cite{Hoang:2010gu}. This approach is 
currently not feasible at NNLO, where no external tools exist to determine 
the remainder terms, and the $\alpha_s$-correction to the coefficient 
functions of the four-electron operators must be determined by the 
methods described in the present paper. Also, the present method 
allows us to consider very loose cuts with $\Lambda\sim m_t$ up to 
the total cross section. On the other hand, assuming that
there is no kinematic or dynamical enhancement in the QCD corrections to 
the remainder contributions, part of the $\alpha_s$-corrections to the 
non-resonant contributions has already been analyzed in~\cite{Hoang:2010gu},
again for intermediate-sized cuts, which in our approach correspond to
NNLO contributions not yet known.

We have checked that the expansion of our
result~(\ref{eq:NLOmatching}) for the
non-resonant contributions in $\Lambda^2/m_t^2\ll 1$ matches the  series
obtained in~\cite{Hoang:2010gu} when $\alpha_s=0$ for the 
$\Gamma_t/\Lambda$ and $\Gamma_t \Lambda/m_t^2$ terms. In this limit, 
diagrams $h_{5}$--$h_{10}$, which are part of the remainder contributions not
calculable within the phase-space matching approach, first contribute 
with $\Gamma_t \Lambda^3/m_t^4$ terms.

\subsection{Comparison to the \boldmath $e^+e^- \to W^+W^- b\bar{b}$ Born cross section from MadGraph}
\label{sec:Madgraph}

Before discussing the final result for the $W^+W^- b\bar{b}$ cross 
section in the top anti-top resonance region including Coulomb 
resummation, we compare the EFT prediction without strong-interaction
effects to the full-theory (Standard Model) $W^+W^- b\bar{b}$ Born cross 
section computed with MadGraph/MadEvent~\cite{Alwall:2007st}.
The input parameters are chosen to be
\begin{equation}
M_Z = 91.1876\,{\rm GeV} \;,\qquad M_W = 80.398\,{\rm GeV}\;,\qquad M_H=120\,{\rm GeV}\,,
\nonumber
\end{equation}
\begin{equation}
G_\mu =1.166367\times 10^{-5}\,{\rm GeV}{}^{-2} \;,\qquad
m_t =172.0\,{\rm GeV} \;,\qquad
V_{tb}=1 \,,
\label{eq:input}
\end{equation}
whereas the on-shell Weinberg angle $c_w=M_W/M_Z$ and the fine-structure 
constant in the $G_\mu$-scheme, $\alpha\equiv \sqrt{2}G_\mu M_W^2 s_w^2/\pi$, 
are derived quantities. Tree-level amplitudes generated by MadGraph
implement the fixed-width prescription for the top-quark propagator.
For the numerical value of $\Gamma_t$ we use the Born 
formula~(\ref{eq:Gammatop}) in both MadGraph and the EFT predictions, 
thus neglecting $\alpha_s$-corrections to the top-width value. 
MadGraph uses a finite bottom-quark mass whereas we neglect it in the EFT
calculation.

\begin{figure}[!tb]
\begin{center}
\includegraphics[width=0.95\textwidth]{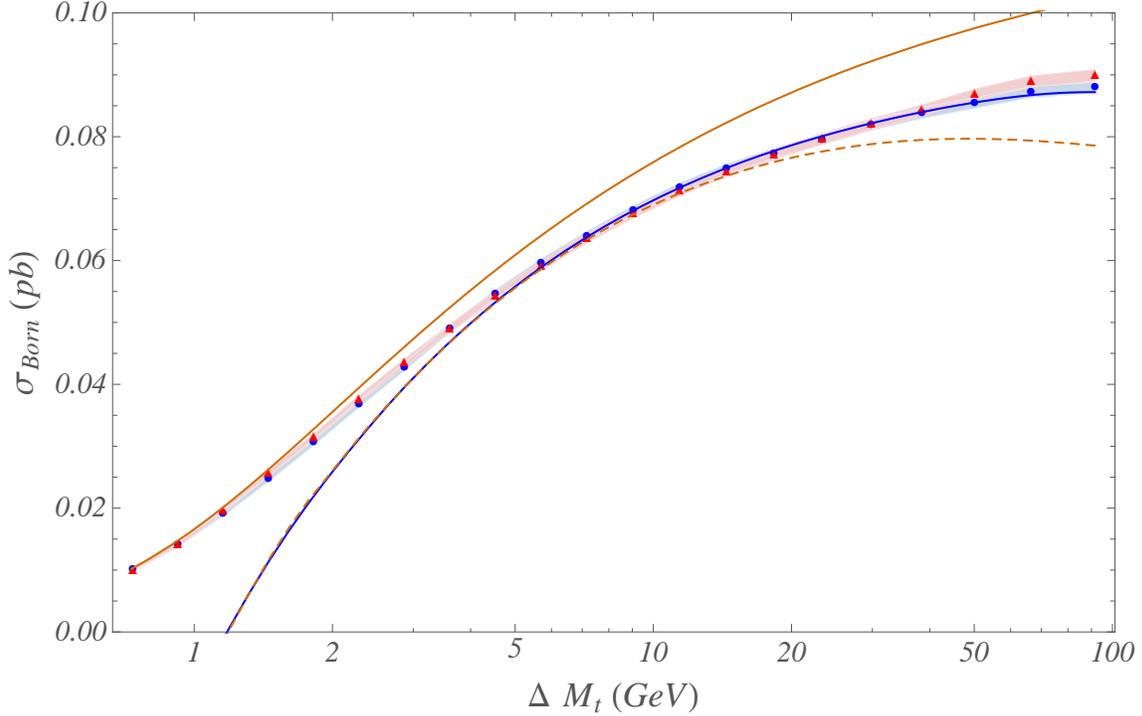}
\caption{Invariant-mass cut dependence of
the $e^+e^-\to W^+W^-b\bar{b}$ Born cross section at $s=4m_t^2$.
The $\Delta M_t$ values range from $\Delta M_t = 0.734\,$GeV 
corresponding to $\Lambda^2=m_t \Gamma_t$ to
the kinematic bound, $\Delta M_{t,\rm max}=m_t-M_W=91.602\,$GeV.
The (red) triangle-shaped dots represent the Standard-Model
cross section computed with MadGraph; for the (blue) circular dots
we have omitted the diagrams with internal Higgs lines. The 
corresponding shaded bands show the 
statistical uncertainty of the MadEvent integration.
Lower solid (blue) line: NNLO EFT approximation given 
in~(\ref{eq:sigmaalphas0}).
Upper solid (brown) line: LO EFT tight-cut prediction (\ref{eq:sigmatight}).
Dashed (brown) line: $\alpha_s=0$ NNLO results with invariant-mass
cuts from~\cite{Hoang:2010gu} (not including the remainder contributions). 
The LO Born EFT result, which does not depend on $\Delta M_t$, is not shown 
in the Figure. It equals $\sigma_{t \bar t,\, \alpha_s=0}^{(0)}=0.1139$~pb 
at $s=4m_t^2$.}
\label{fig3}
\end{center}
\end{figure}

In Figure~\ref{fig3} we analyze the invariant-mass cut dependence of
the $e^+e^-\to W^+W^-b\bar{b}$ Born cross section calculated at $s=4m_t^2$.
The values shown there range from the $\Delta M_t$ value
corresponding to $\Lambda^2=m_t \Gamma_t$ up to
the maximum value allowed by the kinematics, $\Delta M_{t,\rm max}=m_t-M_W$.
The (red) triangle-shaped dots represent the Standard-Model
$e^+e^-\to W^+W^-b\bar{b}$ Born cross section computed with MadGraph, 
while for the (blue) circular dots
we have omitted  in the MadGraph amplitude generation the diagrams 
with internal Higgs lines. The shaded bands represent the 
statistical uncertainties on the MadEvent
integration, calculated as the cross section divided by the square root of
the number of selected events that pass the invariant-mass cut. The small
difference between both sets of points for large values of $\Delta M_t$
is explained below. The lower solid (blue) line is the NNLO
EFT approximation to the full Born result, given by
\begin{equation}
\sigma_{\rm Born} = \sigma_{t \bar t,\, \alpha_s=0}^{(0)} 
+ \sigma_{\text{non-res}}^{(1)}
+\sigma_{v^2}^{(2)}
+\sigma_{P\text{-wave}}^{(2)}
+\sigma_{\rm bil}^{(2)}
+\sigma_{\rm abs}^{(2)}\,.
\label{eq:sigmaalphas0}
\end{equation}
The first term on the right-hand side 
is the LO cross section~(\ref{eq:sigmaLO}), with
the QCD radiative corrections switched off ({\it i.e.} setting $\alpha_s=0$).
At $s=4m_t^2$ it amounts to $\sigma_{t \bar t,\, \alpha_s=0}^{(0)}=0.1139$~pb.
Loose cuts on the $bW^+$ and $\bar{b}W^-$ invariant masses contribute first
at the NLO in the EFT power counting through the non-resonant contributions
$\sigma_{\text{non-res}}^{(1)}$ discussed in Section~\ref{sec:nonresonant}.
The $v^2$-suppressed $S$-wave current describing the top anti-top pair
production in the effective theory gives the NNLO correction 
$\sigma_{v^2}^{(2)}$. Using the non-relativistic equation of motion for 
the top quark it can be taken into account by replacing 
$G_{\rm C}^{(0)}(0,0;{\cal E}) \to (1-{\cal E}/(3m_t))\,
G_{\rm C}^{(0)}(0,0;{\cal E})$
in~(\ref{eq:sigmaLO}).  The term $\sigma_{P\text{-wave}}^{(2)}$
is generated by the matrix element of the production (decay)
$P$-wave operators, which are $v$-suppressed with respect the leading
$S$-wave operators~(\ref{LPlead}). 
It reads
\begin{equation}
\sigma_{P\text{-wave}}^{(2)} 
= \left[\big(C_{p,P\rm \text{-wave}}^{(v)} \big)^2+ 
\big( C_{p,P\rm \text{-wave}}^{(a)} \big)^2\right] 
\frac{4N_c}{3 m_t^2} 
\,\mbox{Im} \, G_{P\text{-wave}}^{(2)}
\,,
\label{eq:sigmaPwave}
\end{equation}
with
\begin{eqnarray}
C_{p,P\rm \text{-wave}}^{(v)} &=& 4\pi\alpha
\, \frac{v_e a_t}{s-M_Z^2} \,,
\nonumber\\
C_{p,P\rm \text{-wave}}^{(a)} &=& -4\pi\alpha
\,\frac{a_e a_t}{s-M_Z^2} \,,
\end{eqnarray}
and 
\begin{equation}
G_{P\text{-wave}}^{(2)} = \frac{m_t^4}{4\pi}
\left(-\frac{{\cal E}}{m_t}\right)^{\!3/2}
,
\end{equation}
the $P$-wave component of the free Green function at zero distance
that can be obtained by setting $\alpha_s=0$ in the
$\overline{\rm MS}$-renormalized expression for the $P$-wave Coulomb Green
function~\cite{Hoang:2001mm}. The term $\sigma_{\rm bil}^{(2)}$ 
originates from the effective-theory matrix
elements with a single insertion of the NNLO corrections 
to the bilinear quark terms in the non-relativistic Lagrangian,
$\delta {\cal L} =1/(8m_t^3) \times \psi_t^\dagger 
( \vec{\partial}^{\,2} + i m_t \Gamma_t)^2\psi_t$, and the corresponding 
anti-quark terms. Written as corrections to the zero-distance 
Green function, it reads
\begin{equation}
\sigma_{\rm bil}^{(2)} 
= \left[{C_p^{(v)}}^2+{C_p^{(a)}}^2\right] 2N_c 
\,\mbox{Im} \, \left( \delta G_{\rm kin}^{(2)}
+\delta G_{\Gamma_t}^{(2)}
+\delta G_{\Gamma_t^2}^{(2)} \right)\,,
\label{eq:sigmakin}
\end{equation}
where $\delta G_{\rm kin}^{(2)}$,
$\delta G_{\Gamma_t}^{(2)}$ and 
$\delta G_{\Gamma_t^2}^{(2)}$ correspond to the insertion
of the $\vec{\partial}^4/(8m_t^3)$, 
$i\Gamma_t\vec{\partial}^{\,2}/(4m_t^2)$
and $-\Gamma_t^2/(8m_t)$ operators, respectively, and read
\begin{align}
\delta G_{\rm kin}^{(2)} &= 
\frac{5m_t^2}{32\pi}\,\left( -\frac{{\cal E}}{m_t} \right)^{3/2} \,,
\\
\delta G_{\Gamma_t}^{(2)} &=  
\frac{3m_t^2}{16\pi}\,i\,\frac{\Gamma_t}{m_t}\,\left( -\frac{{\cal E}}{m_t} 
\right)^{1/2}\,,
\\
\delta G_{\Gamma_t^2}^{(2)} &= -\frac{1}{32\pi}\,
\frac{\Gamma_t^2}{( -{\cal E}/m_t )^{1/2}}\,.
\end{align}
The terms $\delta G_{\Gamma_t}^{(2)}$ and $\delta G_{\Gamma_t^2}^{(2)}$ 
are pure electroweak contributions. While the former
corresponds to the lifetime-dilatation correction studied in
\cite{Hoang:2004tg}, $\delta G_{\Gamma_t^2}$
has not been considered so far among the NNLO electroweak corrections. 
Numerically, the latter term is rather small. Note that 
$\delta G_{\Gamma_t}^{(2)}$ can be obtained by replacing 
$m_t\to m_t -i \,\Gamma_t/2$ in~(\ref{eq:sigmaLO}) and expanding to 
first order in the correction; similarly, 
$\delta G_{\Gamma_t^2}^{(2)}$ follows from replacing 
$\Gamma_t\to \Gamma_t + i \,\Gamma_t^2/(4 m_t)$ in~(\ref{eq:sigmaLO}) and 
the first-order expansion in the correction. In both cases, only 
the term with $\alpha_s=0$ is needed for comparison with the 
Born cross section.
The last term in~(\ref{eq:sigmaalphas0}), $\sigma_{\rm abs}^{(2)}$,
arises from the absorptive parts in the matching coefficients of the 
production operators ${\cal O}_p^{(v)}$ and ${\cal O}_p^{(a)}$ that 
were determined in~\cite{Hoang:2004tg}. It reads
\begin{equation}
\sigma_{\rm abs}^{(2)} 
= 2\left[ C_p^{(v)}C_p^{(v),\rm abs}+ C_p^{(a)}C_p^{(a),\rm abs}\right] 2N_c 
\,\mbox{Re} \, G^{(0)}_0\,,
\end{equation}
with $G^{(0)}_0$ the free Green function. The explicit expressions
for the coefficients $C_p^{(v),\rm abs}$ and $C_p^{(a),\rm abs}$
can be found in~\cite{Hoang:2004tg} (there denoted 
$C_V^{bW,\rm abs}$ and $C_A^{bW,\rm abs}$, respectively).\footnote{The
imaginary part of the field-rescaling factor called $\varpi$ in 
\cite{Beneke:2004km,Beneke:2007zg} is taken into account in the 
calculation of these coefficients.} 
The term $\sigma_{\rm abs}^{(2)}$ reproduces the
interference of single-resonant 
and double-resonant $e^+e^-\to W^+W^- b\bar{b}$ diagrams in the
kinematic region where both $bW^+$ and $\bar{b}W^-$ have invariant
masses of order $m_t$. Within the threshold 
expansion~\cite{Beneke:1997zp}, these contributions
arise from precisely the cut diagrams $h_1$--$h_4$ in Figure~\ref{fig1},
but with one hard and one potential loop, the latter corresponding to the
loop containing the top and anti-top propagators. Let us recall that for 
the non-resonant contributions 
the two loop momenta have to be taken as hard, so the absorptive parts in
$C_p^{(v),\rm abs}$ and $C_p^{(a),\rm abs}$ and
the non-resonant contributions come from different loop-momentum regions and 
there is no double-counting. The additional $v$-suppression of the 
hard-potential with respect to the hard-hard contributions from diagrams 
$h_1$--$h_4$ is explained by the different power counting of the 
loop momentum in the hard ($r_0\sim \vec{r}\sim m_t$) and potential 
($r_0\sim \vec{r}^{\,2}/m_t\sim m_t v^2$) regions.
The sum of all the NNLO corrections in~(\ref{eq:sigmaalphas0})
gives a positive shift of 4.1~fb for $s=4m_t^2$.

Already for values of the invariant-mass cut 
$\Delta M_t \gtrsim 5\,{\rm GeV}\approx 3.5\,\Gamma_t$ we find an 
excellent agreement between the full-theory Born cross section computed with 
MadGraph and the
effective-theory prediction for loose cuts.
To account for invariant-mass cuts
in the range $\Delta M_t \lesssim \Gamma_t$ we need to implement
the tight-cut prescription in the effective-theory calculation.
This is done in the upper solid curve of Figure~\ref{fig3},
which is obtained by introducing the proper step functions 
into the one-loop cut integral of the LO effective-theory result.
Precisely, the LO tight-cut prediction reads
\begin{equation}
\sigma_{t\bar{t},\rm tight}^{(0)} 
= \left[{C_p^{(v)}}^2+{C_p^{(a)}}^2\right] 2N_c 
\,\mbox{Im} \, G_{0,\rm tight }^{(0)} \,,
\label{eq:sigmatight}
\end{equation}
with
\begin{eqnarray}
\mbox{Im} \, G_{0,\rm tight }^{(0)} & = &\frac{\Gamma_t^2}{2}
\int \frac{d^4r}{(2\pi)^4}
\frac{ \theta(\Delta M_t +\frac{E}{2}+r_0-\frac{\vec{r}^{\,2}}{2m_t})\,
\theta(\Delta M_t +\frac{E}{2}-r_0-\frac{\vec{r}^{\,2}}{2m_t})}
{ \left[ (\frac{E}{2}+r_0-\frac{\vec{r}^{\,2}}{2m_t})^2+
\frac{\Gamma_t^2}{4}\right]
\left[ ( \frac{E}{2}-r_0-\frac{\vec{r}^{\,2}}{2m_t})^2+
\frac{\Gamma_t^2}{4}\right] } \nonumber \\[3mm]
& = &\frac{m_t \Gamma_t}{4\pi^3}
\int_0^{\frac{2\Delta M_t}{m_t} }\,\frac{dy}{y} \, 
\Big( y+\frac{E}{m_t} \Big)^{\!1/2} \,  
\bigg( \, \frac{1}{ y - i \frac{\Gamma_t}{m_t} }
  \arctan \left( \frac{\frac{2\Delta M_t}{m_t}- y}
{\frac{\Gamma_t}{m_t} + i y} \right) 
+ {\rm h.c.} \bigg) \,,\qquad
\label{eq:ImGtight}
\end{eqnarray}
which we evaluate at $E=0$ for the result displayed in Figure~\ref{fig3}. 
We see that this expression agrees well with the full Born cross section 
for very small $\Delta M_t$. The agreement could be extended to larger 
values by implementing the cuts in the NNLO terms, but we do not 
pursue this here.

Finally, the dashed curve corresponds to the $\alpha_s=0$ NNLO formula 
with invariant-mass cuts obtained in~\cite{Hoang:2010gu}
(not including the remainder contributions), which contains 
the same NNLO pieces as our result~(\ref{eq:sigmaalphas0}) 
but for $\delta G_{\Gamma_t^2}$. As mentioned before, the pieces 
depending on the invariant-mass cut in the formula in~\cite{Hoang:2010gu} 
match the leading and subleading terms of the expansion in 
$\Lambda^2/m_t^2$ of our NLO non-resonant contributions, and the two 
results should therefore agree well when $\Delta M_t\ll \Delta M_{t,\rm max}$. 
Figure~\ref{fig3} demonstrates this and also shows that as expected, 
the difference between our result (solid blue line) 
and the formula from~\cite{Hoang:2010gu} increases with $\Delta M_t$: 
at $\Delta M_t=35$~GeV it amounts to about $4$~fb, while for the total 
cross section ($\Delta M_{t,\rm max}$)
it reaches almost 9~fb, which roughly agrees with the size of the 
single-top non-resonant contributions, that are not accounted for in 
the phase-space matching approach.

As mentioned above, Figure~\ref{fig3} shows two sets of MadGraph points.
The red triangles represent the complete tree-level cross section. They agree
with our effective-theory prediction for loose cuts $\Delta M_t \gtrsim
5$~GeV, but for $\Delta M_t \gtrsim 40$~GeV and for the
total cross section they rise to values of up to 3~fb above our result.
This small, but visible difference is mostly due to diagrams with
intermediate Higgs lines which account for 2~fb of this difference. 
Eliminating the Higgs diagrams from the MadGraph amplitude gives the blue
dots which agree with the effective-theory result also for large $\Delta M_t$
within the statistical uncertainty of the MadEvent integration.
The Higgs contribution originates
almost exclusively from phase-space regions where the $b\bar{b}$ invariant
mass is close to the Higgs-mass value $M_H=120$~GeV,
\emph{i.e.} from on-shell Higgs bosons decaying into a $b\bar{b}$ pair.
Despite the smallness of the bottom Yukawa coupling, such diagrams can
yield sizable contributions because the Higgs width is small and the
branching fraction of $H \to b\bar{b}$ is large (about 70\%).
The distribution of the Higgs contributions with respect to the invariant
masses of the $b W^+$ or $\bar{b} W^-$ pairs is relatively broad, while the
same invariant-mass distributions of the contributions with (anti)top lines
are sharply peaked around the top mass. This explains why the Higgs
contributions are only visible for very loose or no cuts.
Such small contributions from Higgs diagrams constitute a reducible background
which can easily be eliminated by cuts on the $b\bar{b}$ invariant mass
when one is only interested in the $t\bar{t}$ resonance region.
This justifies that we have not considered contributions from nearly
on-shell Higgs bosons in our effective-theory approach although they
formally count as NLO contributions as explained in
Section~\ref{NLOsummary}.
\begin{figure}[!p]
\begin{center}
\includegraphics[width=0.59\textwidth]{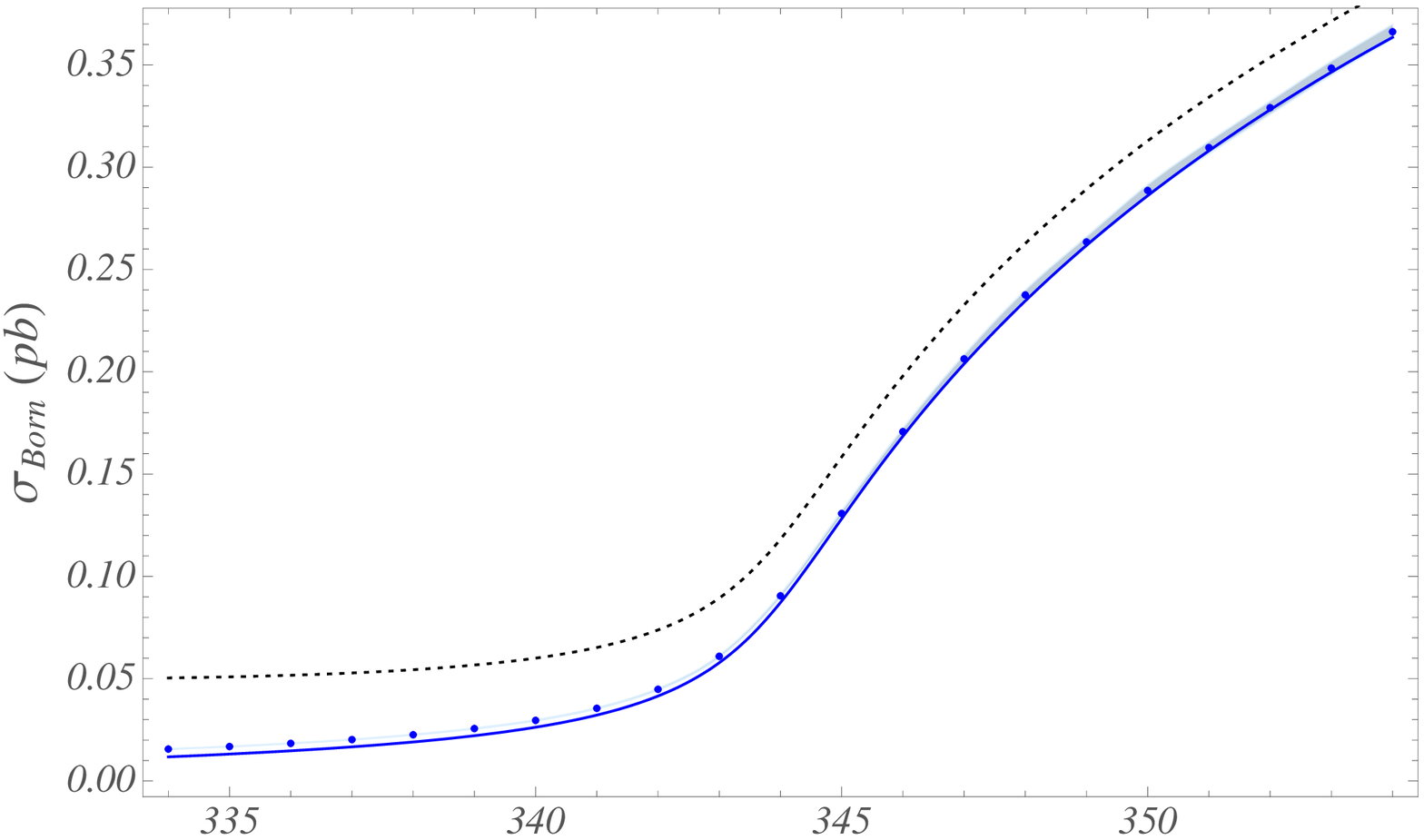}\\[1mm]
\includegraphics[width=0.59\textwidth]{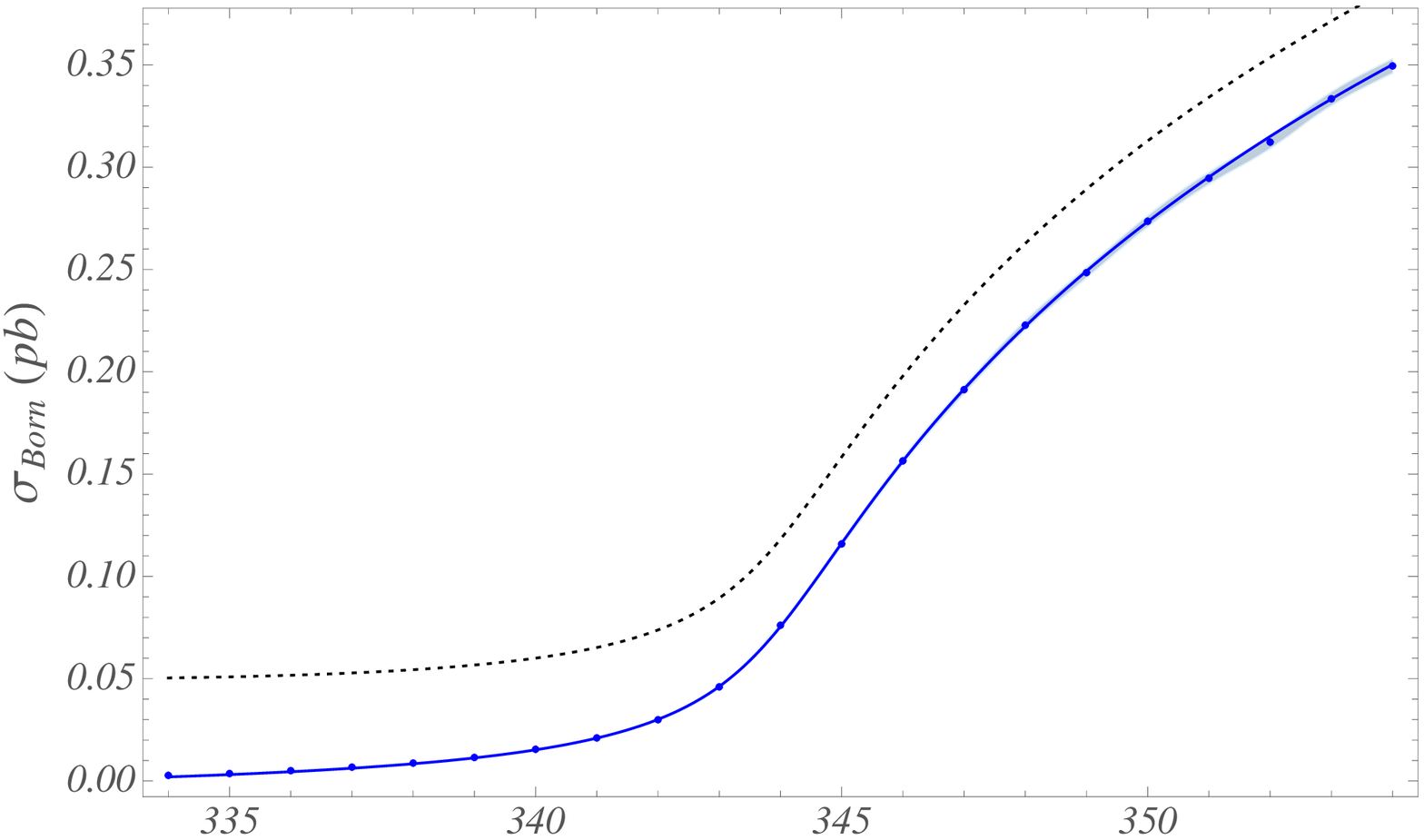}\\[1mm]
\includegraphics[width=0.59\textwidth]{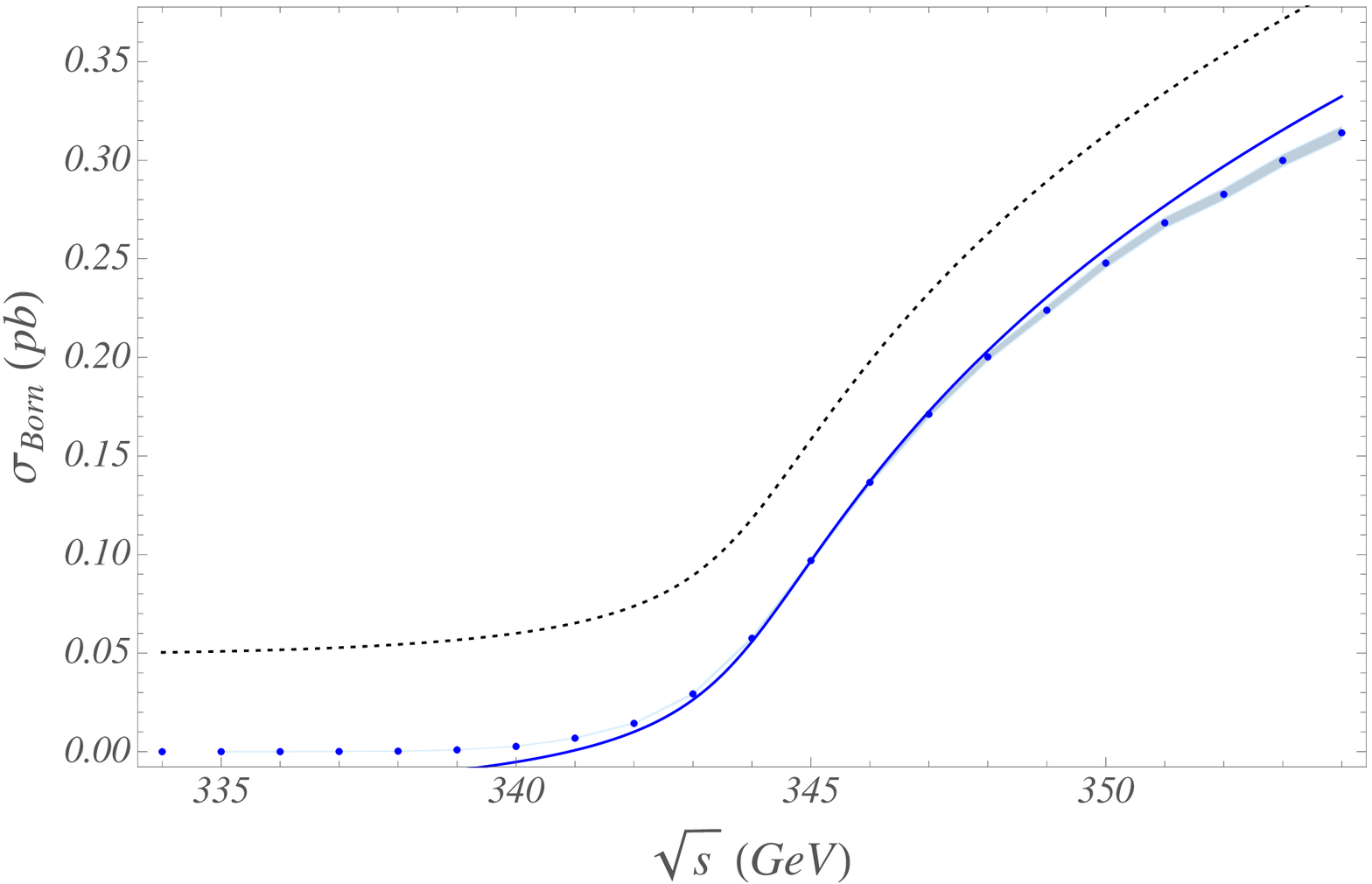}
\caption{NNLO EFT approximation to the Born $W^+W^-b\bar{b}$ cross 
section with (solid blue line) and without (dotted black line) the 
NLO non-resonant contributions as a function of the centre-of-mass energy. 
Upper, middle and lower panels correspond to 
$\Delta M_t=(\Delta M_{t,\rm max},\,15,\,5)$~GeV.
The (blue) dots are the full-theory result
computed with MadGraph, and the shaded band is the 
statistical uncertainty of the MadEvent integration. }
\label{fig4}
\end{center}
\end{figure}

The energy dependence of the effective-theory
prediction~(\ref{eq:sigmaalphas0}) for the total cross section (upper panel),
$\Delta M_t=15$~GeV (middle panel) and $\Delta M_t=5$~GeV (lower panel)
is shown in Figure~\ref{fig4} in the interval $\sqrt{s}=(334,354)$~GeV.
The blue dots correspond to the full-theory Born $W^+W^-b\bar{b}$ 
cross section, while the solid (blue) line is the 
corresponding prediction at NNLO from the unstable-particle effective-theory.
For the total cross section the tiny difference between both results
is mainly due to the Higgs contributions explained above. 
The agreement improves for smaller values of $\Delta M_t$, as seen 
in the middle panel of Figure~\ref{fig4}.
For $\Delta M_t=5$~GeV the effective-theory expansion quickly degrades
as we move away from the threshold. This is evident below threshold,
where the effective-theory result becomes negative.
The large negative shift of about 24--39~fb for the total cross section
given  by the NLO non-resonant contributions is clearly seen in 
Figure~\ref{fig4} once we draw the NNLO effective-theory result without
the non-resonant contributions (dotted line). The shift becomes larger 
as we tighten the invariant-mass cut, up to 38--48~fb for 
$\Delta M_t=15$~GeV.

\section{Final results}
\label{sec:results}

\begin{figure}[!tb]
  \begin{center}
  \includegraphics[width=0.8\textwidth]{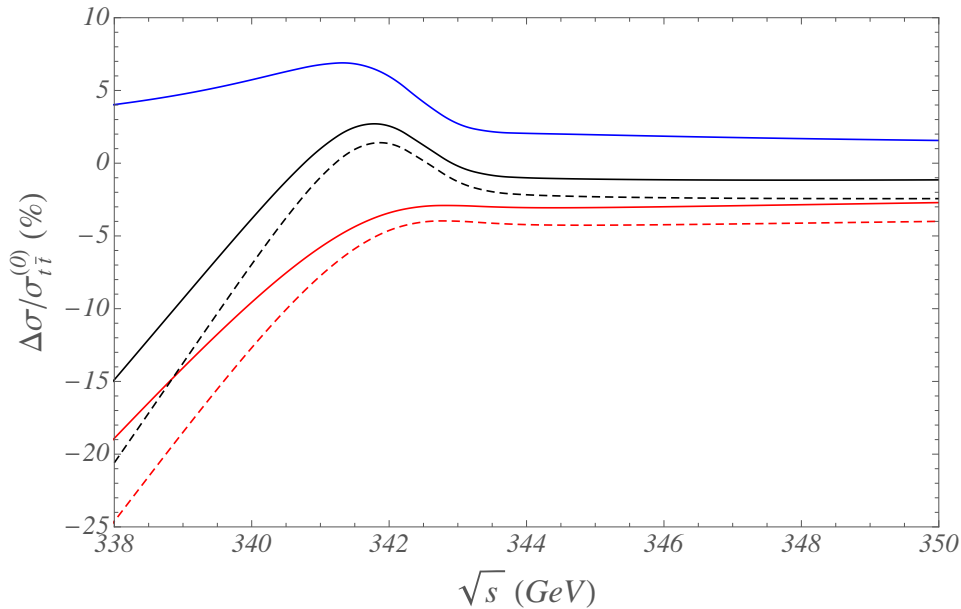}
  \caption{Relative sizes of the QED and non-resonant corrections with
    respect to the $t\bar{t}$ LO cross 
section in percent: $\sigma^{(1)}_{\rm QED}/\sigma^{(0)}_{t\bar t}$ (upper solid blue line),
$\sigma^{(1)}_{\text{non-res}}/\sigma^{(0)}_{t\bar t}$ for 
the total cross section 
(lower solid red line)
and $\Delta M_t=15$~GeV (lower dashed red line). The relative size
of the sum of the QED and non-resonant corrections is represented by the 
middle (black) lines,
for $\Delta M_{t,\rm max}$ (solid) and $\Delta M_t=15$~GeV (dashed).
}
  \label{fig5}
  \end{center}
\end{figure}

In this section we compare the size of the NLO non-resonant and QED 
corrections to the LO effective-theory approximation~(\ref{eq:sigmaLO}) 
for the $e^+e^-\to W^+W^-b\bar{b}$ cross section. This now includes 
the summation of Coulomb corrections proportional to $(\alpha_s/v)^n$ 
to all orders in the strong coupling, which produces the characteristic 
peak structure in the top anti-top resonance region.
As mentioned in Section~\ref{NLOsummary}, we do not discuss the pure 
NLO QCD corrections, which have already been studied in the literature. 
Moreover it is known that the NLO QCD corrections are rather large 
(and negative) and cancelled to a large extent by the NNLO QCD 
corrections~\cite{Hoang:1998xf,Melnikov:1998pr,Beneke:1999qg,Hoang:1999zc,Yakovlev:1998ke,Nagano:1999nw,Penin:1998mx}. Hence
the conclusions that can be drawn by comparing the electroweak corrections 
with the NLO QCD corrections may not be valid once higher-order QCD 
corrections are considered. For the results in this section we choose 
 $\alpha_s(30\,{\rm GeV})=0.142$ for the value of 
the QCD coupling that enters in the LO Coulomb Green function.
The relative sizes of the NLO electroweak corrections with respect 
to the LO result are displayed in Figure~\ref{fig5}.

The upper (blue) solid  line shows the ratio
$\sigma^{(1)}_{\rm QED}/\sigma^{(0)}_{t\bar t}$ (in percent),
where the QED correction $\sigma^{(1)}_{\rm QED}$ is obtained as
\begin{equation}
\sigma^{(1)}_{\rm QED}= \Big(
\sigma^{(0)}_{t\bar t}|_{\alpha_s C_F \to \alpha_s C_F+\alpha Q_t^2} \Big)
-\sigma^{(0)}_{t\bar t}\,.
\label{eq:sigmaQED}
\end{equation}
The QED contribution represents a correction of about 2\% above threshold 
and rises to a maximum of 7\% just below the peak.
The lower (red) solid line is obtained as the ratio 
$\sigma^{(1)}_{\text{non-res}}/\sigma^{(0)}_{t\bar t}$  (in percent) 
with $\Delta M_t$ set by the kinematic bound. The non-resonant contributions
give a constant negative shift of about 3\% above threshold. Below 
threshold the relative size of the non-resonant corrections is very large, 
since the LO result rapidly vanishes, reaching up to 19\%. Hence below 
threshold they represent the leading electroweak correction
to the total $t\bar{t}$ cross section. The sum of both NLO electroweak 
corrections compared to the LO cross section is shown in the middle solid 
(black) line. We observe a partial cancellation of the QED and
non-resonant corrections in the peak region and at energies above.
A sensitivity to the invariant-mass cut $\Delta M_t$ in the $bW^+$ and 
$\bar{b}W^-$ subsystem enters first at NLO through the non-resonant 
contributions. Restricting the available phase-space for the final-state 
particles by tightening the invariant-mass cuts $\Delta M_t$
makes the non-resonant contributions even more important. This is shown by 
the dashed lines in Figure~\ref{fig5}, corresponding to 
$\Delta M_t=15$~GeV, for the relative size of the non-resonant
correction (lower dashed line) and of the sum of the two electroweak 
corrections (middle dashed line).

\begin{figure}[!tb]
  \begin{center}
  \includegraphics[width=0.73\textwidth]{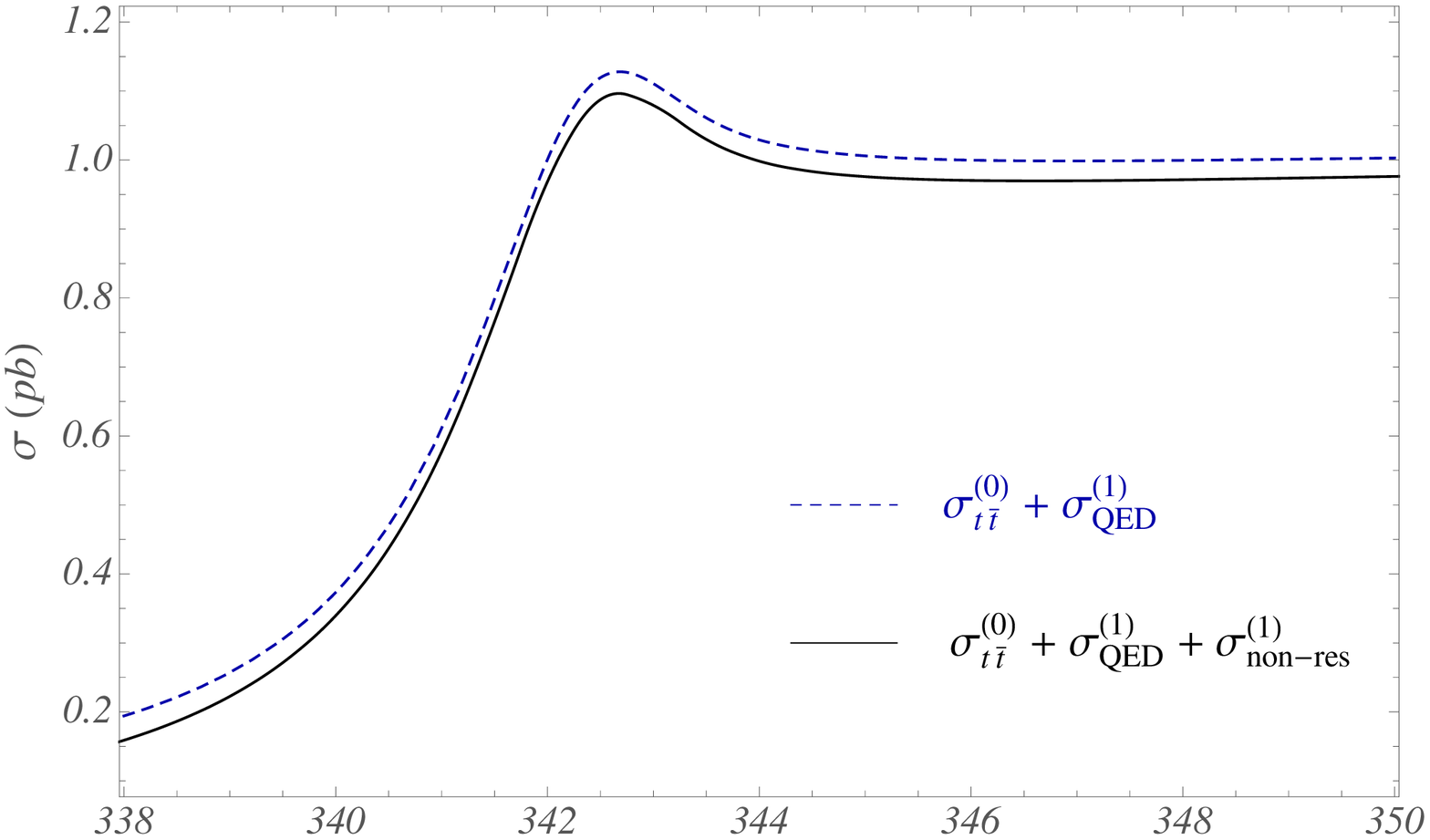}\\[5mm]
  \includegraphics[width=0.73\textwidth]{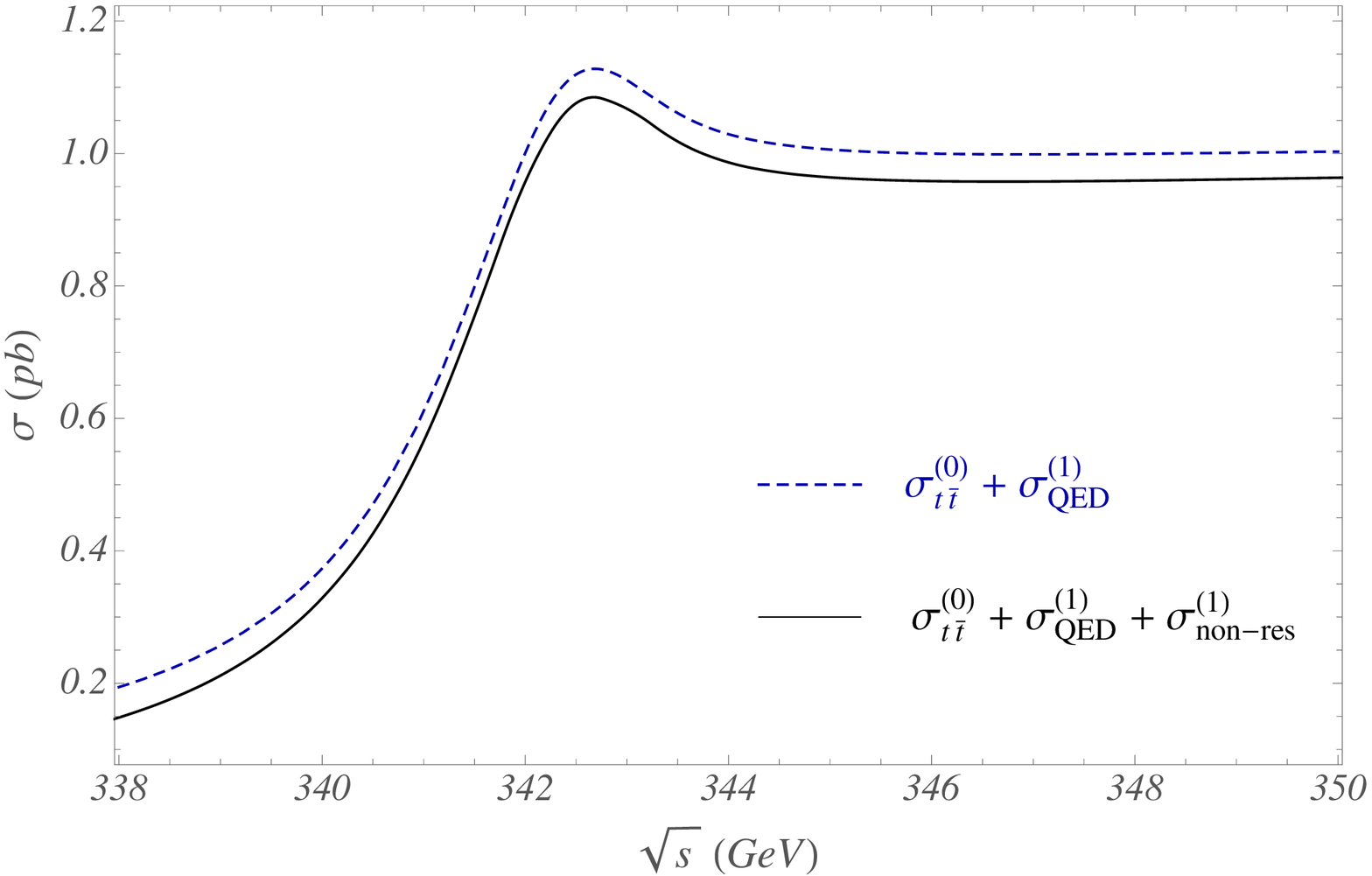}
  \caption{$e^+e^-\to W^+W^-b\bar{b}$ cross section with LO QCD effects and NLO electroweak
corrections at energies close to threshold. The dashed (blue) lines represent $\sigma^{(0)}_{t\bar t}+\sigma^{(1)}_{\rm QED}$ 
and the solid lines
$\sigma^{(0)}_{t\bar t}+\sigma^{(1)}_{\rm QED}+\sigma^{(1)}_{\text{non-res}}$.
The results in the upper panel correspond to the total cross section
($\Delta M_t=\Delta M_{t,\rm max}$), while for the lower panel
$\Delta M_t=15$~GeV.}
  \label{fig6}
  \end{center}
\end{figure}

Aside from the pure QCD corrections, the effective-theory NLO prediction 
for the $e^+e^-\to W^+W^-b\bar{b}$ cross section is displayed by the 
solid lines in Figure~\ref{fig6}. The upper panel corresponds to the total
cross section and for the lower panel $\Delta M_t=15$~GeV. The absolute
size of the non-resonant correction is given by the difference between the 
dashed lines, which only include the QED NLO correction, and the solid ones. 
These negative shifts amount to 27--35~fb for the total cross section and
39--46~fb for $\Delta M_t=15$~GeV for $\sqrt{s}$ in the interval $(338,350)$~GeV.

\section{Summary}
\label{sec:conclusion}

We presented an analysis of the $e^+e^- \to W^+W^-b\bar{b}$ process 
in the top anti-top resonance production region $\sqrt{s}\approx 2 m_t$
extending the methods developed in~\cite{Beneke:2007zg,Actis:2008rb}
for $W$-pair production to the more complicated case of top quarks. 
Our result may be considered as the first complete NLO calculation 
of top-quark pair production near threshold, when the top width 
is accounted for, which necessarily requires to consider the 
final state $W^+W^-b\bar{b}$ for consistency. The result is 
valid for the total cross section and with invariant-mass cuts 
on the $Wb$ systems as long as the cut $\Delta M_t$ is significantly 
larger than the top width. We also included the case of tight 
cuts in the comparison with the full-theory Born cross section.

We find that the non-resonant contributions reduce the cross 
section in the top anti-top resonance peak region by up to $5\%$, 
which is relevant in view of the high-precision top mass and width 
determination anticipated at the ILC. In the energy region below 
the peak the non-resonant correction is significantly larger, since 
the resonant contribution rapidly decreases with energy, while 
the non-resonant 
one is nearly energy-independent. These conclusions agree 
with the recent work~\cite{Hoang:2010gu}. The approach 
pursued in the present paper allows us to extend them to the 
case of the total cross section and rather loose cuts, and it 
also determines the full non-resonant background within the 
effective theory approach rather than using external information. 
The final numerical result turns out to be dominated by the 
off-shell contribution from the double-resonant diagram $h_1$, 
but only due to a large cancellation within classes of other 
diagrams.

Our work provides a step towards predictions including 
consistently the top width in effective theory calculations of 
$t\bar t$ production, more precisely  $W^+W^-b\bar{b}$ production, 
near threshold also in higher orders in the 
expansion in $\alpha_s$, $\alpha_{\rm EW}$ and $\delta\sim v^2$. 
Going to higher orders will eventually be required to 
cancel the finite-width divergences that appear in the NNLO 
and N$^3$LO pure QCD calculations that are already complete 
or near completion.

\vspace*{0.2cm}
\noindent
\subsubsection*{Acknowledgment}
We thank A.H.~Hoang for comments on the manuscript. 
This work is supported by the DFG Sonder\-forschungsbereich/Transregio~9
``Computergest\"utzte Theoretische Teilchenphysik''.
Feynman diagrams have been drawn with the packages 
{\sc Axodraw}~\cite{Vermaseren:1994je} and 
{\sc Jaxo\-draw}~\cite{Binosi:2008ig}. We acknowledge the use of 
the computer programs \textsc{FORM}~\cite{Vermaseren:2000nd} and
\textsc{FeynCalc}~\cite{Mertig:1990an} for 
some parts of the calculation.


\appendix
\section{Formulae for the \boldmath $h_i^X$ functions}
\label{sec:appendix}

We present here integral representations of the functions~$h_i^X(x,y)$
defined in (\ref{eq:hi}) and used in (\ref{eq:NLOmatching}).
Their arguments are $x=M_W^2/m_t^2$ and the cut-dependent
integration limit $y=\Delta^2/m_t^2=1-\Lambda^2/m_t^2$
satisfying $x \le y \le 1$.

\vspace*{0.4cm}\noindent
Diagram $h_1$:
\begin{align}
h_1^V(x,y) &=-\frac{1}{64\pi^2}
\int_{y}^1 dt\, (1-t)^{-3/2}\, \bigg[  \,
\frac{ \Gamma_t(t)}{ \Gamma_t^{\rm Born} }\,
\sqrt{9-t} \Big( 27 + 73 t - 3 t^2 - t^3 \Big) - 192\sqrt{2} \bigg]\,,
\nonumber
\\
h_1^A(x,y) &=-\frac{1}{64\pi^2}
\int_{y}^1 dt\, \sqrt{ \frac{9-t}{1-t} }\, 
\frac{ \Gamma_t(t) }{ \Gamma_t^{\rm Born} }\,
\Big( 27 + 4 t + t^2 \Big) \,,
\nonumber
\\
h_1^{VA}(x,y) &=-\frac{1}{64\pi^2}
\int_{y}^1 dt\, \sqrt{ \frac{9-t}{1-t} }\,  
\frac{ \Gamma_t(t) }{ \Gamma_t^{\rm Born} }\,
\Big( 27 - 2 t - t^2 \Big) \,,
\label{eq:h1int}
\end{align}
%
with the ``off-shell'' top width defined as 
\begin{equation}
\Gamma_t ( t ) = \Gamma_t^{\rm Born}  
\frac{\left( 1- \frac{x }{t} \right)^2 
\left( 1+ \frac{2 x }{t} \right) t}
{( 1- x)^2 ( 1 + 2 x) }\,.
\label{eq:Gammatopoff}
\end{equation}
The last term within the square 
brackets in the first line of~(\ref{eq:h1int}) is a 
subtraction term which regularizes the integrand at $t=1$. 
The integration of minus this subtraction term using dimensional 
regularization yields the terms quoted in the first line on the 
right-hand side of~(\ref{eq:NLOmatching}).

\vspace*{0.4cm}\noindent
Diagram $h_2$:
\begin{align}
h_2^V(x,y) &=
\frac{1}{4 \pi ^2  (x-1)^2 (2 x+1)}
\int_{y}^1 dt\,
\nonumber \\
&
\Bigg\{
\frac{t \left(2 x^2+7 x-5\right)-x \left(2 x^2+15 x+31\right)}{1-t} \,
f_+(t,x)
\nonumber \\
&
+
\frac{t-x}{16  t^2 } \,
\sqrt{\frac{9-t}{1-t} } \,
 \Big(t^4+t^3 (x+4)-t^2 \left(2 x^2+12 x+53\right)-t x (24 x+85)-6 x^2\Big)
\Bigg\}\,,
\label{eq:h2Vint}
\end{align}

\begin{align}
h_2^A(x,y) &=
\frac{1}{4 \pi ^2  (x-1)^2 (2 x+1)}
\int_{y}^1 dt\,
\nonumber \\
&
\Bigg\{
\frac{t \left(2 x^2+5 x-3\right)-x \left(2 x^2+13
   x+17\right) }{ 1-t }\,
f_+(t,x)
\nonumber \\
&
+
\frac{t-x}{16 t^2 } \,
\sqrt{\frac{9-t}{1-t} } \,
 \Big(t^4+t^3 (x+2)-t^2 \left(2 x^2+14 x+27\right)-t x (20 x+59)+6 x^2\Big)
\Bigg\}\,,
\label{eq:h2Aint}
\end{align}
with
\begin{equation}
f_\pm(t,x)=\ln \left(\frac{\sqrt{(9-t)(1-t)} (x-t)-t(t-5) \pm x(t+3)}{\sqrt{(9-t)(1-t)} (t-x)-t(t-5) \pm x(t+3) }\right)
\,.
\label{eq:fint}
\end{equation}
\\
\noindent
Diagram $h_3$:
\begin{align}
h_3^V(x,y) &=
\frac{1}{8 \pi ^2 (x-1)^2 x (2 x+1)}
\int_{y}^1 dt\,
\nonumber \\
&
\Bigg\{
\frac{1}{ 1-t }
\Bigl(t^3 x+t^2 \left(4 x^2-6 x+2\right)+t \left(-2 x^3-29 x^2+2\right)
\nonumber \\
&\qquad \qquad
  +x \left(4 x^3+36 x^2+x-13\right)\Bigr) \,
f_-(t,x)
\nonumber \\
&
+\frac{t-x }{8   t^2 } \, \sqrt{\frac{9-t}{1-t}} \,
\Big (t^4 x+t^3 \left(x^2-4 x-4\right)+t^2 \left(-2 x^3-28 x^2+15 x+4\right)
\nonumber \\
& \qquad \qquad \qquad \qquad 
-t x \left(24 x^2+69 x+92\right)-6 x^3 \Big)
\Bigg\}\,,
\label{eq:h3Vint}
\end{align}

\begin{align}
h_3^{A}(x,y) &= \frac{1}{8 \pi ^2 (x-1)^2 x (2 x+1)}
\int_{y}^1 dt\,
\nonumber \\
&
\Bigg\{
\frac{1}{t-1 }\,
\Bigl(t^3 x+t^2 \left(4 x^2-6 x+2\right)+t \left(-2 x^3-23 x^2+20 x+2\right)
\nonumber \\
&\qquad \qquad
  +x \left(4 x^3+24 x^2-33 x+7\right)\Bigr) \,
f_-(t,x)
\nonumber \\
&
-\frac{t-x}{8  t^2 } \, \sqrt{\frac{9-t}{1-t}} \,
\Big(t^4 x+t^3 (x-6) x+t^2 \left(-2 x^3-30 x^2+53 x+16\right)
\nonumber \\
&\qquad\qquad\qquad\qquad
+t x \left(-20 x^2-75 x+56\right)+6 x^3
\Big)
\Bigg\}\,.
\label{eq:h3Aint}
\end{align}
\\
\noindent
Diagram $h_4$:
\begin{align}
h_4^V(x,y) &=
-\frac{3 x}{64 \pi^2 (1-x)^2 (2 x+1)}
\int_{y}^1 dt\, \sqrt{ \frac{9-t}{1-t} } 
\nonumber \\
&
\Bigg\{
\,
-(t-x) \, \frac{t (2 t x+t-17 x-1)+23 x}{6 t x^2}
\nonumber \\
&
+\frac{t^3 x+t^2 \left(2 x^2-5 x+2\right)+t (x+2)-x (86 x+13)}{3 \sqrt{(9-t)(1-t)} \, x^2} \,
f_-(t,x)
\nonumber \\
&
+\frac{1}{2} \left(t^3-5 t^2-25 t+93\right) g^{(0)}(t,x)
+\frac{t^4-4 t^3-106 t^2+604 t-495}{2 \sqrt{(9-t)(1-t)}} \, g^{(1)}(t,x)
\nonumber \\
&
+\frac{1}{2} \left(-t^3+21 t^2-119 t+99\right) g^{(2)}(t,x)
-\frac{1}{2} \Bigl((9-t)(1-t)\Bigr)^{3/2} \, g^{(3)}(t,x)
\Bigg\}
 \,,
\label{eq:h4Vint}
\end{align}
\begin{align}
h_4^A(x,y) &=
\frac{3 x}{64 \pi^2 (1-x)^2 (2 x+1)}
\int_{y}^1 dt\, \sqrt{ \frac{9-t}{1-t} } 
\nonumber \\
&
\Bigg\{\,
(t-x) \, \frac{ t^2 x-2 t (4 x+1)-7 x }{3 t x^2}
\nonumber \\
&
-\frac{t^3 x+t^2 \left(2 x^2-5 x+2\right)+t (21 x+2)+(7-30 x) x}{3 \sqrt{(9-t)(1-t)} \, x^2} \,
f_-(t,x)
\nonumber \\
&
-\frac{1}{2} (t-3) (t-1)^2 \, g^{(0)}(t,x)
+\frac{-t^4+4 t^3+74 t^2-284 t+207}{2 \sqrt{(9-t)(1-t)}} \, g^{(1)}(t,x)
\nonumber \\
&
+\frac{1}{2} \left(t^3-21 t^2+119 t-99\right) g^{(2)}(t,x)
+\frac{1}{2} \Bigl((9-t)(1-t)\Bigr)^{3/2} \, g^{(3)}(t,x)
\Bigg\}
 \,,
\label{eq:h4Aint}
\end{align}
with
\begin{equation}
g^{(i)}(t,x) =
\frac{1}{2} \int_{-1}^1 dz \, \frac{z^i}{\sqrt{ a  z^2 - b z + c}} \,
\ln\left(\frac{d - e z - k \sqrt{ a z^2 - b z + c }}
  {d - e z + k \sqrt{ a z^2 - b z + c }}\right) 
\label{eq:gint}
\end{equation}
and
\begin{align}
a &=4 (9-t)(1-t) (1-x) \,,\nonumber \\
b &=4 \sqrt{(9-t)(1-t)} \, (x-2) (t+2 x-3) \,,
\nonumber \\
c &=t^2 \left(x^2+4\right)-2 t \left(5 x^2-8 x+12\right)+25 x^2-48 x+36 \,,
\nonumber \\
d &=-2 t^2+t \bigl((x-1) x+6\bigr)-x (x+3)\,,
\nonumber \\
e &= -\sqrt{(9-t)(1-t)} \, \bigl(t (x-2)+x\bigr)\,,
\nonumber \\
k &=t-x \,.
\end{align}
\\
\noindent
Diagram $h_5$:

\begin{align}
h_5(x,y) &=
\frac{3 x}{1024 \pi^2 (1-x)^2 (2 x+1)}
\int_{y}^1 dt\, \sqrt{ (9-t)(1-t) } 
\nonumber \\
&
\Bigg\{ \,
-\frac{t-x}{6 t^2 x^3 \bigl(t^2 x+t (4-6 x)+x (4x-3)\bigr)} \,
\Big(t^4 x (8 x+1)+t^3 \left(7 x^2+39 x+4\right)
\nonumber \\
&\qquad
+t^2 \left(-32 x^3+53 x^2+11 x+84\right)+t x \left(28 x^2+13 x-27\right)+3 x^2 (4 x-3)
\Big) 
\nonumber \\
&
+\frac{4 \bigl((t-9) x^2+(t-11) x-6\bigr)}{3 x^3 \sqrt{(9-t)(1-t)} }
 \,
f_-(t,x)
+\frac{4 (t+3)}{x}\,
g^{(0)}(t,x)
\nonumber \\
&
-\frac{4\sqrt{(9-t) (1-t)}}{ x}\,
 g^{(1)}(t,x)
- 4x \left(t^2-8 t+15\right) \, j^{(0)}(t,x)
\nonumber \\
&
-8x (t-4) \sqrt{(9-t)(1-t)} \, j^{(1)}(t,x)
-4x (9-t)(1-t) \, j^{(2)}(t,x)
\nonumber \\
&
+8 \, \frac{t^2 x + t(2-4x)-x-6}{x} \,
  \widetilde{j}^{(0)}(t,x)
+8 \sqrt{(9-t)(1-t)} \, \frac{ t x+2}{ x} \, \widetilde{j}^{(1)}(t,x)
\Bigg\}
 \,,
\label{eq:h5int}
\end{align}
with
\begin{align}
j^{(i)}(t,x) &=
\left( 1- \frac{x}{t} \right) \int_{-1}^1 dz \, z^i \,
 \frac{ (c_3 d_1 +c_1 d_3) (c_2^2+8 c_1 c_3) -6 c_1 c_2 c_3 d_2 }{2 c_1 c_3 (c_2^2-4c_1 c_3)^2 }
\nonumber \\
&
+ \int_{-1}^1 dz \, z^i \,
\frac{ (c_2^2+2c_1 c_3) d_2 -3 c_2 (c_1 d_3+c_3 d_1) }
  { (c_2^2-4c_1 c_3)^2 \, \sqrt{ a  z^2 - b z + c} }
\,
\ln \left( \frac{d - e z - k \sqrt{ a z^2 - b z + c }}
  {d - e z + k \sqrt{ a z^2 - b z + c }} \right)
 \,,
\label{eq:jint}
\end{align}
\begin{align}
\widetilde{j}^{(i)}(t,x) &=
- \left( 1- \frac{x}{t} \right) \int_{-1}^1 dz \, z^i \,
 \frac{2c_1 \widetilde{d}_3-c_2 \widetilde{d}_2 }{c_1 (c_2^2-4c_1 c_3) }
\nonumber \\
&
- \int_{-1}^1 dz \, z^i \,
\frac{ 2 c_3 \widetilde{d}_2 - c_2 \widetilde{d}_3 }{ c_2^2-4c_1 c_3 } \,
\frac{1}{\sqrt{ a  z^2 - b z + c}} \,
\ln \left( \frac{d - e z - k \sqrt{ a z^2 - b z + c }}
  {d - e z + k \sqrt{ a z^2 - b z + c }} \right)
\,,
\label{eq:jtildeint}
\end{align}
and
\begin{align}
c_1 &=
\frac{x \bigl((t-6) t+4 x-3\bigr)}{t}+4
\,,
\nonumber \\
c_2 &=
\frac{4 \widetilde{z} \bigl(t (x-2)+x\bigr)-t^2 (x-4)-t x (x+3)+x^2}{t}
\,,
\nonumber \\
c_3 &=
\frac{x (2 \widetilde{z}-t) (2 \widetilde{z}-x)}{t}
\,,
\nonumber \\
\widetilde{z} &=\frac{1}{4} \left(-\sqrt{(9-t) (1-t)} \, z+t+3\right)
\,,
\nonumber \\
d_1 &=
\frac{(t-7) (t-3) t^2+\bigl(t (t+2)+9\bigr) x^2+\bigl(t (t+2)-27\bigr) t x}{t^2}
\,,
\nonumber \\
d_2 &=
\frac{4 \widetilde{z} \bigl((t-3) t^2+(t+3) x^2+(t-3) t x\bigr)
  -t \bigl(t^2 (3 x+4)+t x (3 x-23)+13 x^2\bigr)}{t^2}
\,,
\nonumber \\
d_3 &=
\frac{4 \widetilde{z}^2 \left(t^2+t x+x^2\right)-6 \widetilde{z} t x (t+x)+3 t^2 x^2}{t^2} \,,
\nonumber \\
\widetilde{d}_2 &=
-\frac{t (t+x-5)+3 x}{2 t} \;,
\qquad
\widetilde{d}_3 =
x-\frac{\widetilde{z} (t+x)}{t}
\,.
\end{align}
\\
\noindent
Diagram $h_6$:
\begin{align}
h_6(x,y) &=-\frac{1}{128 \pi^2 (x-1)^2 x^2 (2 x+1)}
\int_{y}^1 dt\, \sqrt{ (9-t)(1-t) } 
\nonumber \\*
&
\Bigg\{ \,
\frac{t-x}{t^2 \bigl(t^2 x+t (4-6 x)+x (4 x-3)\bigr)} 
\Big( t^4 x \left(6 x^2+7 x+1\right)
\nonumber \\
&\qquad
+t^3 \left(-21 x^3+9 x^2+35 x+4\right)+t^2 \left(-40 x^4-224 x^3+36 x^2-21 x+84\right)
\nonumber \\
&\qquad
+t x \left(60 x^3-33 x^2+25
   x-27\right)+3 x^2 (4 x-3)
\Big)
\nonumber \\
&
+\frac{2  \bigl( \left(t^2-8 t+19\right) x^2+2 (t+1) x^3-2 (t-17) x+24\bigr)}
  {\sqrt{(9-t)(1-t)}} \,
f_-(t,x)
\nonumber \\
&
-6 (t+3) x^2 (3 x+2)\,
g^{(0)}(t,x)
+
6 \sqrt{(9-t)(1-t)} \, x^2 \bigl((t+6) x+2\bigr)
\, g^{(1)}(t,x)
\nonumber \\
&
-6 (9-t)(1-t) x^3 \, g^{(2)}(t,x)
\nonumber \\
&
-6 x^2 \bigl(t^2 x (x+4)-2 t \left(5 x^2+8 x-4\right)+33 x^2+20 x-24\bigr) \,
  \widetilde{j}^{(0)}(t,x)
\nonumber \\
&
+12 \sqrt{(9-t)(1-t)} \, x^2 \bigl(-2 (t-2) x+7 x^2-4\bigr)
 \, \widetilde{j}^{(1)}(t,x)
\nonumber \\
&
-6 (9-t)(1-t) x^4
\, \widetilde{j}^{(2)}(t,x)
\Bigg\}
 \,.
\label{eq:h6int}
\end{align}
\\
\noindent
Diagram $h_7$:
\begin{align}
h_7(x,y) &=
\frac{3 x}{64 \pi^2 (1-x)^2 (2 x+1)}
\int_{y}^1 dt\, \frac{ \sqrt{(9-t)(1-t)} }{5-t} 
\nonumber \\
&
\Bigg\{ \,
-\frac{ (t-5) (t-x) (2 t x+t-15 x)}{6 t x^2}
\nonumber \\
&
+\frac{t^3 x-t^2 \left(2 x^2+9 x-2\right)+3 t \left(8 x^2+3 x-2\right)+x \left(16 x^2-18 x-17\right)}{3 \sqrt{(9-t)(1-t)} \, x^2} \,
f_-(t,x)
\nonumber \\
&
- \frac{4 \bigl(t^2 x^2+t \left(-8 x^2+x-1\right)+\left(-4 x^2+2 x+13\right) x\bigr)}{3 \sqrt{(9-t)(1-t)} \, x^2} \,
f_+(t,x)
\nonumber \\
&
-\frac{1}{2} \bigl(-t^3+t^2 (2 x+13)-5 t (4 x+3)+66 x-45\bigr)\,
g^{(0)}(t,x)
\nonumber \\
&
+\frac{1}{2} \sqrt{(9-t) (1-t)} \, \bigl((t-2) t+28 x-39\bigr)\,
 g^{(1)}(t,x)
\nonumber \\
&
-\frac{1}{2} (9-t)(1-t) (t+2 x-11) \,g^{(2)}(t,x)
-\frac{1}{2} \bigl((9-t) (1-t)\bigr)^{3/2}\,
 g^{(3)}(t,x)
\nonumber \\
&
-\frac{1}{2} \bigl(t^3-t^2 (2 x+7)+t (20 x+3)-66 x-45\bigr)\,
\widetilde{g}^{(0)}(t,x)
\nonumber \\
&
+\frac{1}{2} \sqrt{(9-t) (1-t)} \, \bigl(t (t+4)-7 (4 x+3)\bigr)\,
\widetilde{g}^{(1)}(t,x)
\nonumber \\
&
-\frac{1}{2} (9-t)(1-t) (t-2 x+1)\,
 \widetilde{g}^{(2)}(t,x)
+\frac{1}{2} \bigl((9-t) (1-t)\bigr)^{3/2} \,
  \widetilde{g}^{(3)}(t,x)
\Bigg\}
 \,,
\label{eq:h7int}
\end{align}
with
\begin{equation}
\widetilde{g}^{(i)}(t,x) =
-\frac{1}{2} \int_{-1}^1 dz \, \frac{z^i}
  {\sqrt{ -\widetilde{a}  z^2 - \widetilde{b} z + \widetilde{c}}} \,
\ln \left( \frac{\widetilde{d} - \widetilde{e} z
  - k \sqrt{ -\widetilde{a} z^2 - \widetilde{b} z + \widetilde{c} }}
  {\widetilde{d} - \widetilde{e} z + k \sqrt{ -\widetilde{a} z^2 -
  \widetilde{b} z + \widetilde{c} }} \right) 
\label{eq:gtildeint}
\end{equation}
and
\begin{align}
\widetilde{a} &=
-\frac{1}{4} (9-t)(1-t) \left(t^2-2 t-16 x+1\right) 
\,,
\nonumber \\
\widetilde{b} &=\frac{1}{2} \sqrt{(9-t) (1-t)} \, \bigl((t-6) t-8 x-3\bigr)
  (t-2 x-1) 
\,,
\nonumber \\
\widetilde{c} &= 
  \frac{t^4-4 t^3 (x+3)+t^2 \left(4 x^2+44 x+30\right)
  -4 t \left(10 x^2+35 x-9\right)+100 x^2+228 x+9}{4}\,,
\nonumber \\
\widetilde{d} &= \frac{1}{2} \bigl(-t^3+3 t^2 (x+2)+t \left(-2 x^2-16 x+3\right)+x (2 x+21)\bigr)\,,
\nonumber \\
\widetilde{e} &= \frac{1}{2} \sqrt{(9-t) (1-t)} \, \bigl(t (-t+x+1)+7 x\bigr)\,.
\end{align}
\\
\noindent
Diagram $h_8$:
\begin{align}
h_8(x,y) &=
\frac{1 }{4 \pi ^2 (x-1)^2 (2 x+1)}
\int_{y}^1 dt\,
\Bigg\{
(2 t x+t-2 x-9) \,f_+(t,x)
\nonumber \\
&
+\frac{(t-x)\sqrt{(9-t)(1-t)} }{16  t^2  \bigl( t(1-x) +x (x+3)\bigr)} \,
\Big( t^4 (1-x)-t^3 (x-5)+t^2 \left(3 x^3+49 x^2+44 x-72\right)
\nonumber \\
&\qquad \qquad
-t x \left(2 x^3+23 x^2+107 x+144\right)+6 x^2 \left(x^2+7 x+12\right)
\Big)
\Bigg\}\,.
\label{eq:h8int}
\end{align}
\\
\noindent
Diagram $h_9$:
\begin{align}
h_9(x,y) &=
\frac{1 }{8 \pi ^2  (x-1)^2 x (2 x+1)}
\int_{y}^1 dt\,
\nonumber \\*
&
\Bigg\{
\frac{2\bigl(t \left(2 x^3+7 x^2-7 x+2\right)-x \left(2 x^3+23 x^2+25 x+6\right)\bigr)}
{  t-5 }
\,f_+(t,x)
\nonumber \\
&
+\frac{1}{ t-5}
\Bigl(t^3 x-2 t^2 \left(2 x^2+5 x-1\right)+t \left(-2 x^3+19 x^2+12 x-6\right)
\nonumber \\*
&\qquad \qquad
-x \left(4 x^3+16
   x^2-25 x+17\right)\Bigr)
\,f_-(t,x)
\nonumber \\
&
+\frac{(t-x)\sqrt{(9-t)(1-t)} }{ 8 t^2 } \,
\Big( t^3 x+t^2 \left(x^2-3 x-4\right)+t x \left(-2 x^2+5 x+20\right)
\nonumber \\*
&\qquad \qquad \qquad \qquad \qquad \qquad
+6 x^2 (x+4) \Big)
\Bigg\}\,.
\label{eq:h9int}
\end{align}

\noindent
Diagram $h_{10}$:
\begin{align}
h_{10}(x,y) &=
\frac{1}{4 \pi ^2 (x-1)^2 x^2 (2 x+1)}
\int_{y}^1 dt\,
\nonumber \\
&
\Bigg\{
-\Big( \left(-t^2 +7 t - 3\right) x^2+(4 t+11) x^3+6 x^4-12 x-12 \Big) 
\,f_-(t,x)
\nonumber \\
&
+ \frac{(t-x)\sqrt{(9-t)(1-t)} }{16  t^2  \bigl(t^2 x+t (4-6 x)+x (4 x-3)\bigr)}
\,
\Big(
t^5 x^3+t^4 x \left(x^3+47 x^2+28 x+4\right)
\nonumber \\
&\qquad \qquad \qquad \qquad
+t^3 \left(-2 x^5+19 x^4-221 x^3+64 x^2+124 x+16\right)
\nonumber \\
&\qquad \qquad \qquad \qquad
+t^2 \left(86 x^5+259 x^4-683 x^3+12 x^2-212
   x+336\right)
\nonumber \\ 
&\qquad \qquad \qquad \qquad
+t x \left(-8 x^5+54 x^4+201 x^3-208 x^2+148 x-108\right)
\nonumber \\
&\qquad \qquad \qquad \qquad
+6 x^2 \left(4 x^4+13 x^3-12 x^2+8 x-6\right)
\Big)
\Bigg\}\,.
\label{eq:h10int}
\end{align}

\end{document}